\documentclass[structabstract]{aa}

\usepackage{txfonts}
\usepackage{graphicx}

\usepackage{natbib}

\begin{document}

\title{Profiling filaments: comparing near-infrared extinction and submillimetre data in TMC-1\thanks{{\it Herschel} is an ESA space observatory with science instruments provided by European-led Principal Investigator consortia and with important participation from NASA.}}

\author{
J. Malinen \inst{1} \and
M. Juvela \inst{1} \and
M. G. Rawlings \inst{2,3} \and
D. Ward-Thompson \inst{4} \and
P. Palmeirim \inst{5} \and
Ph. Andr\'e \inst{5}
}

\institute{
Department of Physics, University of Helsinki, P.O. Box 64, FI-00014 Helsinki, Finland; johanna.malinen@helsinki.fi \and
Joint ALMA Observatory / European Southern Observatory, Alonso de C\'{o}rdova 3107, Vitacura 763-0355, Santiago, Chile \and
Joint Astronomy Centre, 660 N. A'ohoku Place, Hilo, HI 96720, U. S. A. \and
School of Physics and Astronomy, Cardiff University, Queen's Buildings, Cardiff CF24 3AA \and
Laboratoire AIM, CEA/DSM-CNRS-Universit\'e Paris Diderot, IRFU/Service d'Astrophysique, C.E.A. Saclay, Orme des Merisiers, 91191 Gif-sur-Yvette, France
}

\date{}

\abstract
{
Interstellar filaments are an important part of the star formation process.
In order to understand the structure and formation of filaments, the filament cross-section profiles are often fitted with the so-called Plummer profile function. Currently this profiling is often approached with submillimetre studies, especially with \emph{Herschel}.
If these data are not available, it would be more convenient if filament properties could be studied using groundbased near-infrared (NIR) observations.
}
{
We compare the filament profiles obtained by NIR extinction and submillimetre observations to find out if reliable profiles can be derived using NIR observations.
}
{
We use J-, H-, and K-band data of a filament north of TMC-1 to derive an extinction map from colour excesses of background stars. We also use 2MASS data of this and another filament in TMC-1. We compare the Plummer profiles obtained from these extinction maps with \emph{Herschel} dust emission maps. We present two new methods to estimate profiles from NIR data: Plummer profile fits to median $A_{\rm V}$ of stars within certain offset 
or directly to the $A_V$ of individual stars. We compare these methods by simulations.
}
{
In simulations the extinction maps and the new methods give correct results to within $\sim$10-20\% for modest densities ($\rho_{\rm c}$ = $10^4$--$10^5$ cm$^{-3}$).
The direct fit to data on individual stars
usually gives more accurate results than the extinction map, and can work in higher density.
In the profile fits to real observations, the values of Plummer parameters are generally similar to within a factor of $\sim$2 (up to a factor of $\sim$5). Although the parameter values can vary significantly, the estimates of filament mass usually remain accurate to within some tens of per cent.
Our results for TMC-1 are in good agreement with earlier results obtained with SCUBA and ISO.
High resolution NIR data give more details, but 2MASS data can be used to estimate approximate profiles.
}
{
NIR extinction maps can be used as an alternative to submm observations to profile filaments. Direct fits of stars can also be a valuable tool in profiling. However, the Plummer profile parameters are not always well constrained, and caution should be taken when making the fits and interpreting the results. 
In the evaluation of the Plummer parameters, one can also make use of the independence of the dust emission and NIR data and the difference in the shapes of the associated confidence regions.
}


\maketitle

\section{Introduction} \label{sect:intro}

Interstellar filaments have long been recognised as an important part of the star formation process. Recent studies have confirmed that filamentary structures are a common feature in dense interstellar clouds and different stages of potentially forming stars are located in these filaments
\citep[e.g.][]{Andre2010,Arzoumanian2011,Juvela2012}.
Therefore, it is important to study the formation and structure of the filaments in order to understand the details of star formation.

The mass structure of molecular clouds can be studied via a number of methods. These include molecular line mapping, observations of dust emission at far-infrared/submillimetre wavelengths, star-counts in the optical and near-infrared wavelengths, and measurements of colour excesses of background stars to produce an extinction map. The surface brightness based on scattered NIR light has also proved to be a potential method for studying mass distributions~\citep[see e.g.][]{Padoan2006,Juvela2006,Juvela2008}.
However, all techniques have their own drawbacks~\citep[see, e.g.,][for a comparison of several methods]{Goodman2009}. For example, line and continuum emission maps are subject to abundance variations (gas and dust, respectively) and variations in the physical conditions. Mass estimates based on dust emission and on the estimation of colour temperature can also be biased because of line-of-sight temperature variations, especially in potentially star forming high density clouds~\citep{Malinen2011}.

The details of filaments have been studied for a long time using various methods, including extinction maps~\citep[see e.g.][and references therein]{Schmalzl2010}.
However, the current interest in studying filament profiles originates from submillimetre observations~\citep[e.g.][]{Nutter2008,Arzoumanian2011,Hill2011,Juvela2012}, especially the high resolution data of \emph{Herschel} Space Observatory~\citep{Pilbratt2010}. Many recent studies have fitted the filament cross-section profiles with the so-called Plummer profile function (see details in Section~\ref{sect:profiles}), because the parameters of the function can give information about the formation and structure of the filament \citep[see more details e.g. in][submitted]{Arzoumanian2011,Juvela2012a}. Large magneto-hydrodynamic (MHD) simulations have shown that filaments can be formed as a natural result of interstellar turbulence~\citep[e.g.][]{Padoan2001}. \citet[][submitted]{Juvela2012a} have studied the observational effects in determining filament profiles from synthetic submillimetre observations.

The use of far-infrared/submillimetre wavelengths is more expensive, as spaceborne satellites are needed. However, in the future groundbased large-area surveys made with submillimetre wavelengths (for example with SCUBA-II) will also become available~\citep[see e.g.][]{Ward-Thompson2007}.
If these data are not available, it would be more convenient if groundbased NIR observations could be used to study the detailed properties of cloud filaments.
The use of dust extinction could also be more reliable, as it is not dependent on the easily biased estimates of colour temperature.
It would also be useful to compare mass structures obtained using different methods. 
We therefore aim to compare the filament Plummer profiles obtained from NIR extinction data to those obtained from submillimetre dust emission observations.

The Taurus molecular cloud is one of the closest (distance $\sim$ 140 pc) and most studied star-forming regions~\citep[see e.g.][and references therein]{Nutter2008,Schmalzl2010,Lombardi2010}. 
The cloud has also been mapped with \emph{Herschel} as part of the Gould Belt Survey~\citep{Andre2010}, see~\citet{Kirk2012} and Palmeirim et al.~(in prep.).
It is therefore a good candidate for studying the structure of filaments.
In this paper, we will examine in detail two cloud filaments in Taurus, comparing the results obtained from NIR and submillimetre data. Additionally, we examine two more distant clouds mapped by \emph{Herschel}.

The contents of this article are as follows: Observations and data reduction are described in Sect.~\ref{sect:observations} and the methods to derive column densities in Sect.~\ref{sect:columndensity}. We present our methods to derive filament profiles in Sect.~\ref{sect:profiles}. The results of profiling observed filaments are shown in Sect.~\ref{sect:results}. We examine the performance of the analysis methods in Sect.~\ref{sect:simulations}, with the help of simulations.
We discuss the methods and results in Sect.~\ref{sect:discussion} and present our conclusions in Sect.~\ref{sect:conclusions}.

\section{Observations and data reduction} \label{sect:observations}

\begin{figure}
\includegraphics[width=8cm]{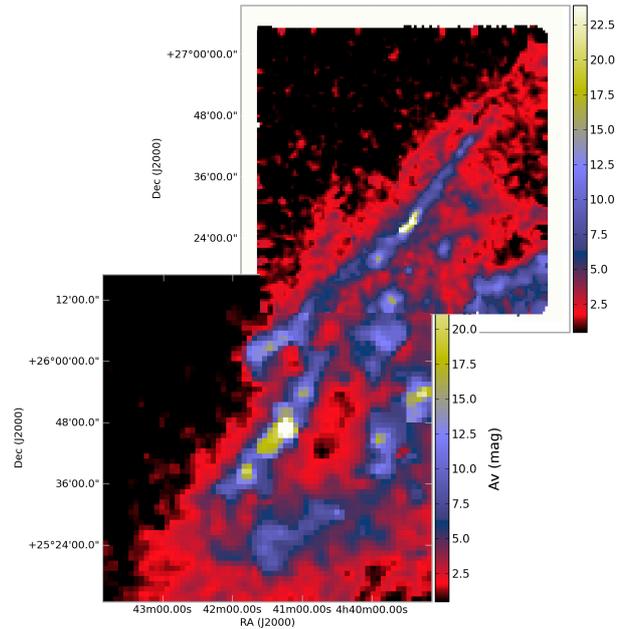}
\caption{Extinction maps of TMC-1 (southern field, 2MASS data) and TMC-1N (northern field, WFCAM data).}
\label{fig:extinctionmap}
\end{figure}

We study two filaments in the Taurus molecular cloud, TMC-1 (also known as Bull's tail \citep{Nutter2008}) and a filament north of TMC-1, which we call TMC-1N. The central coordinates are RA~(J2000) 4h41m30s and Dec~(J2000) +25$^{\circ}$45$'$0$''$ for TMC-1 and RA~(J2000) 4h39m36s and Dec~(J2000) +26$^{\circ}$39$'$32$''$ for TMC-1N. A combined extinction map of the two fields is shown in Fig.~\ref{fig:extinctionmap}.
According to~\citet{Rebull2011}, there are no known young stellar objects (YSOs) in TMC-1N. 
There are several YSOs in TMC-1, however, six of them in the filament analysed here.
YSOs are redder than background stars and could therefore bias the extinction estimates, especially if only a few stars are seen in the middle of the filament. We test the effect of YSOs by comparing results obtained before and after removing them.

We also study two other filaments mapped with \emph{Herschel}, already presented in~\citet{Juvela2012}.
The profiles of these filaments are shown in Appendix~\ref{sect:appendix_A}.

\subsection{WFCAM}

We have used the Wide Field CAMera (WFCAM) \citep{Casali2007} of the United Kingdom InfraRed Telescope (UKIRT) to sample a $1^{\circ} \times 1^{\circ}$ field in the near-infrared (NIR) J-, H-, and K-bands (1.2, 1.6, and 2.2 $\mu m$, respectively) in the Taurus molecular cloud complex field TMC-1N.
Our target field choice was based on the Taurus extinction maps of \citet{Cambresy1999} and \citet{Padoan2002}.

We used the ON-OFF method in order to be able to measure diffuse emission.
The observations were made during 13 nights between 2006--2008 on 4 $\times$ 4 adjacent ON-fields.

The integration times of the ON-fields were 8.5, 11.3, and 27.3 minutes in the J, H, and K band, respectively.
The coordinates of the 4 OFF-fields are (RA~(J2000), Dec~(J2000)): 3h52m and +23$^{\circ}$43$'$, 3h52m and +24$^{\circ}$9$'$, 3h54m and +23$^{\circ}$43$'$, 3h54m and +24$^{\circ}$9$'$.

The data were reduced in the normal pipeline routine\footnote{\tt http://casu.ast.cam.ac.uk/surveys-projects/wfcam/technical}.
In order to study the surface brightness, standard decurtaining methods were needed to remove the stripes caused by the instrument.
The reduced data were converted to 2MASS magnitude scale using the formulas of \citet{Hewett2006} and calibrated using 2MASS stars. The Iraf Daophot-package was used for photometry and Montage\footnote{\tt http://montage.ipac.caltech.edu/} for combining the images.
The median magnitudes for which 
magnitude uncertainty $\sigma \sim0.1^m$
of our WFCAM data are  $J = 20.0^{\rm m}$, $H = 19.1^{\rm m}$, and $K = 18.8^{\rm m}$. These are $4-4.5^{\rm m}$ deeper than 2MASS observations. In the analysis, we used only sources with $\sigma < 0.1^{\rm m}$.
Saturated stars were replaced with corresponding 2MASS stars.
Based on the classification scheme of~\citet{Cantiello2005}, potential background galaxies were removed from our star list, in a manner similar to that of~\citet{Schmalzl2010}.

\subsection{2MASS}

We used Two Micron All Sky Survey (2MASS) (Skrutskie et al. 2006) data to make extinction maps of the fields for which we did not have deeper observations. We also used 2MASS data for TMC-1N in order to perform a comparison with the WFCAM data that contains a significantly larger number of stars.
The sizes of our fields are approximately $1^{\circ} \times 1^{\circ}$. The limiting magnitudes of 2MASS are $J$ = 15.8$^{\rm m}$, $H$ = 15.1$^{\rm m}$, and $K_S$ = 14.3$^{\rm m}$.

\subsection{\emph{Herschel}}

Both TMC-1 and TMC-1N have also been mapped with \emph{Herschel}~\citep{Pilbratt2010} as part of the Gould Belt Survey~\citep{Andre2010}. We used SPIRE~\citep{Griffin2010} 250\,$\mu$m, 350\,$\mu$m, and 500\,$\mu$m maps of both of the fields. 
We obtained the data from the \emph{Herschel} Gould Belt Survey consortium.
The observation identifiers of the data are 1342202252 and 1342202253.

Fields G300.86-9.00 (PCC550) and G163.82-8.44 have been mapped with \emph{Herschel} as part of the Cold Cores project and are described in more detail in~\citet{Juvela2012}.

\section{The column density maps} \label{sect:columndensity}

\subsection{Column densities from the NIR reddening}

The extinction maps were derived from the near-infrared reddening of
the background stars, using 2MASS observations for TMC-1 and
the WFCAM observations for the northern field TMC-1N. For comparison,
an extinction map of TMC-1N was also made using 2MASS data. The extinction values
were calculated with the NICER method \citep{Lombardi2001}, assuming
an extinction curve corresponding to $R_{\rm V}=4.0$ \citep{Cardelli1989}.

For comparison with the N(H$_2$) values below, the extinction values
were converted to N(H$_2$), assuming a relation 
$N({\rm H_2}) \sim$0.94$\times 10^{21}$\,$A_{\rm V}$ \citep{Bohlin1978}. 
The relation itself is derived for diffuse regions, but serves as a good
point of reference for the comparison between the measurements of dust
reddening and dust emission.
The resolution of the derived column density map is 2$\arcmin$ for 2MASS data and 1$\arcmin$ for WFCAM data.

\subsection{Column densities from dust emission}

The maps of the dust colour temperature, $T_{\rm C}$, were calculated
using the SPIRE 250\,$\mu$m, 350\,$\mu$m, and 500\,$\mu$m maps. The
data were convolved to a common resolution of 40$\arcsec$ and the
$T_{\rm C}$ values were estimated pixel by pixel assuming a constant
value of the dust spectral index, $\beta=2.0$. The least-squares fit
of modified black body curves $B_{\nu}(T_{\rm C})\nu^{\beta}$ provides
both the colour temperature and the intensity that were converted to
estimates of column density. The column density was obtained from the
formula
\begin{equation}
   N({\rm H}_2) = \frac{ I_{\nu} }{ B_{\nu}(T) \kappa \mu m_{\rm H}},
\end{equation}
where the dust opacity per total mass, $\kappa$, was assumed to follow
the law 0.1\,cm$^2$/g\,($\nu$/1000\,GHz)$^{\beta}$ that should be
applicable to high density environments \citep{Hildebrand1983,Beckwith1990}.
The resolution of the column density maps derived from \emph{Herschel} observations is 40$\arcsec$.

\section{Modelling the filament profiles} \label{sect:profiles}

We selected the most prominent filament from each field.
The filaments were originally traced by eye and the paths of the filaments were
consequently adjusted to follow the crest of the column density as
derived from the observations of the dust emission. The filament profiles perpendicular to the filament are extracted at small intervals and
the final profile
of the filament is obtained by averaging all data along the filament.

We fit the filament column density profiles $N(r)$ with Plummer-like profiles
\begin{equation}
\rho_{p}(r) = \frac{\rho_{c}}{[1+(r/R_{\rm flat})^2]^{p/2}}
\Rightarrow
N(r)  = A_{p} \frac{\rho_{c}R_{\rm flat}}{[1+(r/R_{\rm flat})^2]^{(p-1)/2}}
\end{equation}
\citep[see e.g.][]{Nutter2008, Arzoumanian2011}. The equation includes the
central density $\rho_{\rm c}$, the size of the flat inner part
$R_{\rm flat}$, and the parameter $p$ that describes the steepness of
the profile in the outer parts.
We include in the fits two additional
parameters to describe a linear background.  
According to~\citet{Ostriker1964}, the value of $p$ should be 4 in the case of an isothermal cylinder in hydrostatic equilibrium.
However, the values that are derived from
observations are often smaller and typically around $p\sim$2
\citep[e.g.][]{Arzoumanian2011,Juvela2012}.
The factor $A_{p}$ is obtained from the formula
$A_{p} = \frac{1}{cos\, i} \int_{-\infty}^{\infty}
\frac{du}{(1+u^2)^{p/2}}$,
where we assume an inclination angle of $i=0$.
In addition to the fitted Plummer profile parameters, we also derive the mass per unit length $M_{\rm line}$ of the filament by integrating column density over the profile.

We use four different methods for extracting the profiles, i.e., \emph{profiling}.
The methods are described in more detail in the following subsections. In short, the profiles are based on a fit to
\begin{flushleft}
\begin{enumerate}
\item a column density map derived from (submillimetre) dust emission map [Method A].
\item a column density map derived from NIR extinction map [Method B].
\item median $A_{\rm V}$ of stars within a certain offset from the filament centre [Method C].
\item $A_{\rm V}$ of individual stars [Method D].
\end{enumerate}
\end{flushleft}
Method $A$ is currently often used for estimating filament properties. The aim of our study is to investigate, if the alternative methods $B$, $C$, and $D$ based on NIR data could also be used for extracting filament profiles.

\subsection{Extraction of profiles from column density maps}

Methods $A$ and $B$ both use a column density map for fitting, but in method $A$, the column density map is derived from a (submillimetre) dust emission map, and in method $B$, from a NIR extinction map. Otherwise, the process for extracting the profile parameters is the same.
We step along the filament with a step size equal to half of the full width at half maximum (FWHM) of the input map. At each position, we construct the local profile that consists of the values read from the map at intervals of 0.3$\times$FWHM along a line perpendicular to the filament length.
Before averaging, the profiles are aligned in order that their peaks match on the offset axis.

Because the observations have a finite resolution, the data are fitted
with a Plummer profile that is convolved to the resolution of the
column density map used. Thus, the obtained parameters of the Plummer
function should correspond to those of the true filament. In addition
to the convolved Plummer profile, we include in the fits a linear
background to take into account the possible extended
background. The final free parameter is the shift of the Plummer
profile that allows for some errors in the position of the originally
selected filament. For the dust emission, this already
follows the maximum column density. However, it is better to allow a shift in the fit. This applies
particularly to the analysis of the extinction where, in principle, the
exact position of the observed filament could differ from the
position deduced from the dust emission (as is the case in TMC-1, see Figs.~\ref{fig:a_f} and \ref{fig:a_f_c}).

The estimated FWHM of the filament profile and mass per unit length, $M_{\rm line}$, depend on how wide a region is used in the fitting and how the background is removed, as there are no clear limits of where the filaments end, particularly in crowded areas. The fitted Plummer parameters, however, should not depend on the width of the fitting region, as long as the region is wide enough to cover also the edges of the filament.
In the fits to the average profile, we only use data where offsets from
the filament centre correspond to distances [-0.3\,pc, +0.3\,pc] for TMC-1 and TMC-1N. For the other filaments shown in Appendix~\ref{sect:appendix_A}, we use bigger offset limits, 0.5\,pc, in order to make sure that these wider and more distant filaments are properly fitted.
We calculate FWHM values for each individual profile and for the median profile after subtracting the minimum. This is directly the observed value without deconvolution with the instrument beam.

\subsection{Extraction of profiles using $A_{\rm V}$ of stars directly}

The previous extinction maps represent the extinction values of the
individual stars that are averaged with a relatively large Gaussian
beam. This is necessary in order that no holes remain in the $A_{\rm
V}$ map and each pixel value corresponds to the average of at least of
a few stars. However, because we are mainly interested in the average
properties of the filaments, we can average all the data along the
filament length and this should allow us to reach a higher resolution
for the profile function. We use two different methods for extracting profiles: median $A_{\rm V}$ values for each offset bin (method $C$) or direct fit to the $A_{\rm V}$ values of individual stars (method $D$). In both of these methods we extract profiles only using stars that are within a given offset from the filament centre: [-0.3\,pc, +0.3\,pc] for TMC-1 and TMC-1N, and [-0.5\,pc, +0.5\,pc] for the other filaments shown in Appendix~\ref{sect:appendix_A}, as in the case of using column density maps.

In method $C$, we calculate the offset from the
filament centre for each star. The stars are binned in narrow offset intervals
(i.e., bins extended along the whole filament) and we calculate the
median $A_{\rm V}$ for each offset with a step size 0.3 times the
bin width. These profiles can again be fitted, taking into account that the binning effectively introduces some loss of resolution.
In the following results and simulations, the width of the offset bins
is always one arc minute, although we have tested also other values.
The drawback of this method is that we cannot
use the average and the dispersion of the $A_{\rm V}$ values of the
neighbouring stars to filter outliers as effectively. 
We calculate the FWHM of the median profile after subtracting the minimum. As with column density maps, this is directly the observed value without deconvolution with the instrument beam.

Method $D$ relies directly on the measurements of the
individual stars, without any spatial averaging. Each star represents
extinction within a very narrow region and the final resolution of the
profile function depends only on the density of the stars along the offset axis.
The influence of each star is weighted not
only according to the uncertainty of its photometry but also inversely
proportionally to the local stellar density around the star. Thus, the
final profile is weighted according to the
area rather than the number of stars within that area. The data are fitted directly
with the Plummer profile. Because the individual stars provide a
practically infinite resolution, no convolution is needed.
In connection with this method we do not calculate any separate FWHM estimates.

\section{Results} \label{sect:results}

\subsection{Analysis of the column density maps}

\begin{figure*}
\includegraphics[width=16cm]{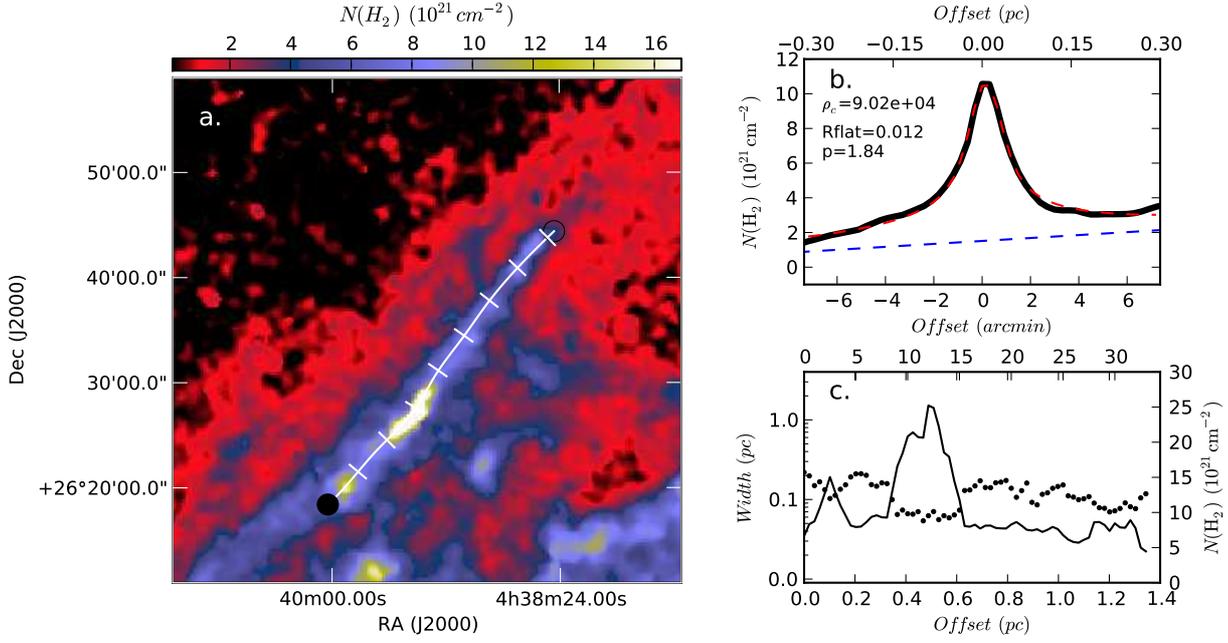}
\caption{TMC-1N column density map and filament profile derived from WFCAM extinction map. a) Column density map. The part of filament used in the fitting is marked with white line (based on the \emph{Herschel} column density map in Fig.~\ref{fig:n_a_f_c}). b) Average column density profile of the filament (black line), fitted Plummer profile (red dashed line) and the baseline of the fit (blue dashed line). Values for fitted Plummer parameters $\rho_{\rm c}$, $R_{\rm flat}$, and $p$ are marked in the figure. c) FWHM values (black circles and left hand scale) and column density along the ridge of the filament (solid line and the right hand scale).}
\label{fig:n_a_f}
\end{figure*}

\begin{figure*}
\includegraphics[width=16cm]{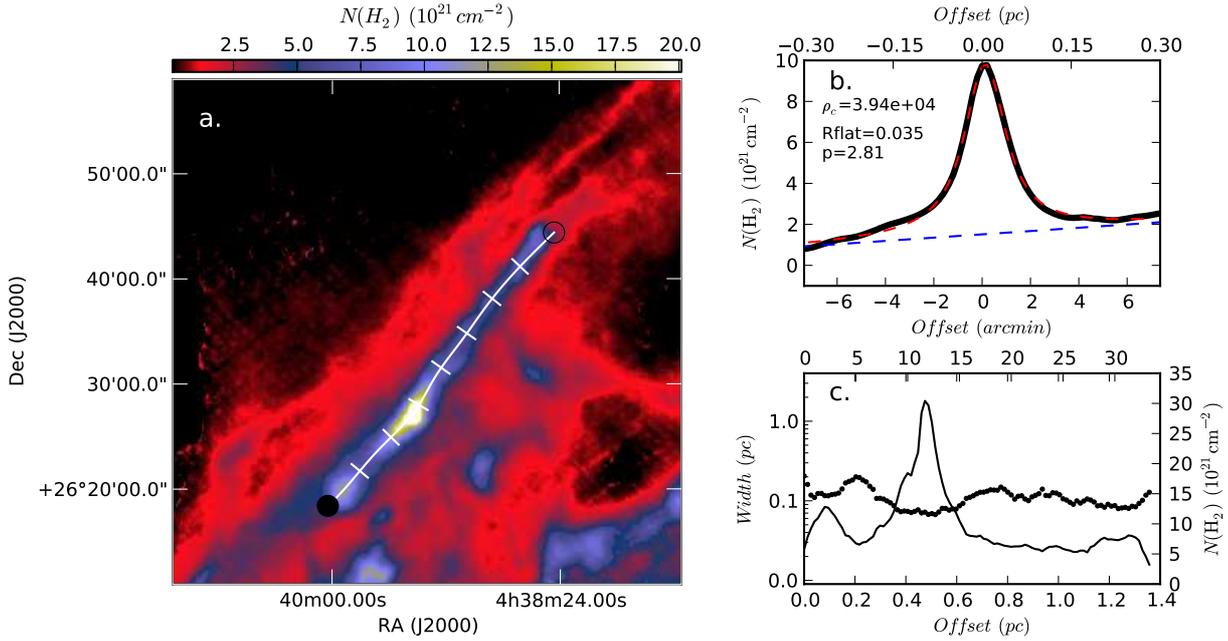}
\caption{TMC-1N column density map and filament profile derived from \emph{Herschel} emission map. See Fig.~\ref{fig:n_a_f} for explanations of the used notation.}
\label{fig:n_a_f_c}
\end{figure*}

Column density maps and filament profiles derived from the WFCAM extinction map and \emph{Herschel} emission map of TMC-1N are shown in Figs.~\ref{fig:n_a_f} and \ref{fig:n_a_f_c}. The filament is traced based on the \emph{Herschel} map and marked with white line in both column density maps. The filament position obtained from the \emph{Herschel} map fits very well also in the WFCAM map. In the \emph{Herschel} map the marked part forms the densest continuous filament, which we call here the full filament. However, in the WFCAM extinction map, the filament continues seamlessly southwards outside the frame.

The profiles obtained by using these two methods ($A$ and $B$) are very similar. The FWHM of the filament is approximately 0.1 pc, which is in agreement with the general results of~\citet{Arzoumanian2011} on filaments in the IC 5146 molecular cloud.
When plotting together the column density and FWHM values along the ridge of the filament, we can see a notable anticorrelation between these parameters. In the densest parts of the filament FWHM gets smaller, possibly indicating that gravity is pulling together the mass and forming narrower condensations in the filament, similarly to the results of~\citet{Juvela2012}.

The column density structure along the ridge of the filament is quite similar in both data sets. In the \emph{Herschel} data, the highest peak is narrower and reaches a column density of $\sim30\cdot10^{21}$ cm$^{-2}$ while in the WFCAM data the maximum is $\sim25\cdot10^{21}$ cm$^{-2}$.
Considering the many uncertainties in the column density estimates (e.g., dust opacity), the values are practically identical. Conversely, the similarity of the column density values suggests that the assumptions used in the column density calculation were reasonable.

\begin{figure*}
\includegraphics[width=16cm]{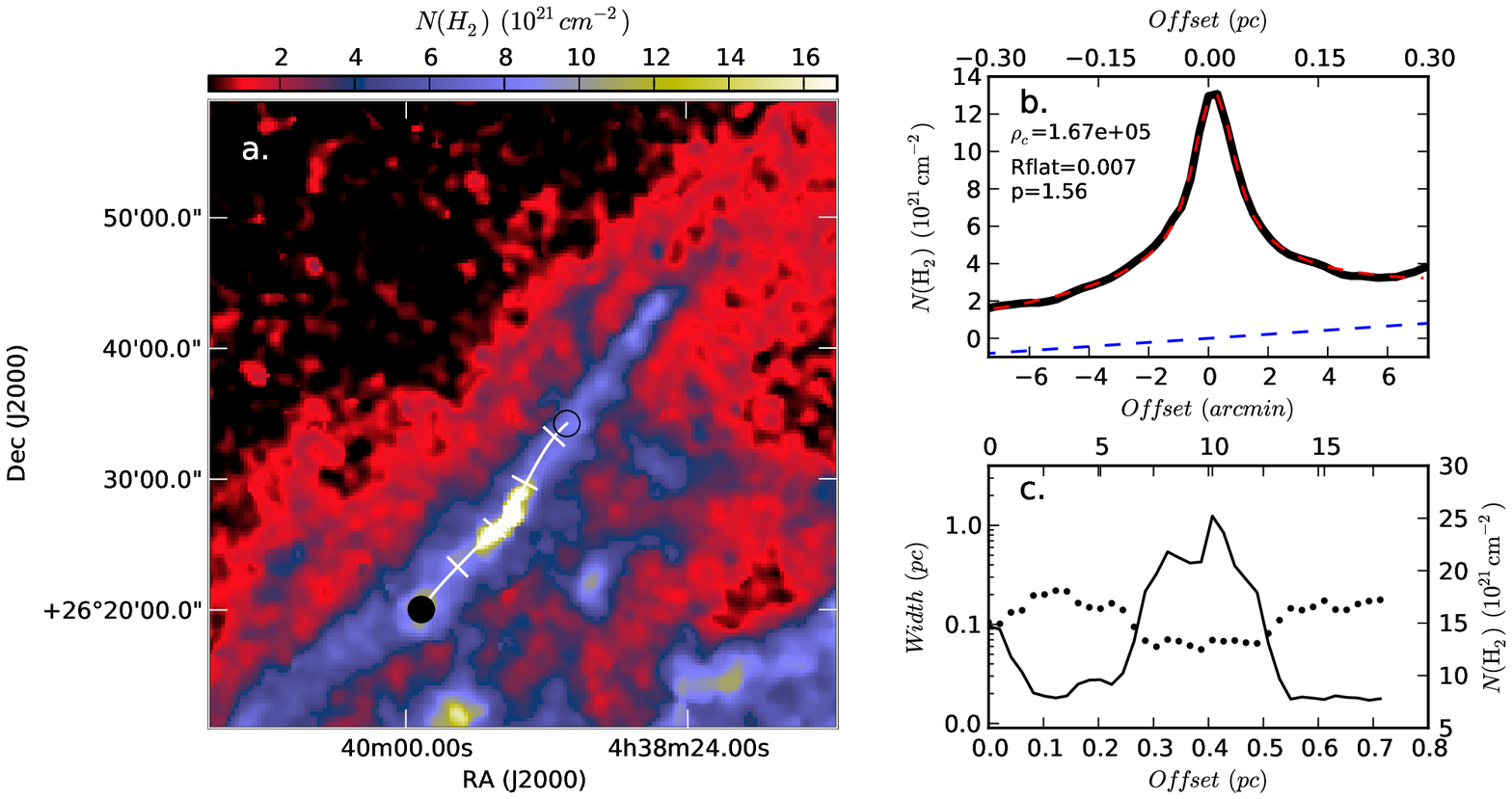}
\caption{TMC-1N column density map and profile of the densest middle part of the filament derived from WFCAM extinction map. See Fig.~\ref{fig:n_a_f} for explanations of the used notation.}
\label{fig:n_a_m}
\end{figure*}

To study the effect of choosing the filament end points, we also extract profiles from the middle part of the filament in TMC-1N, shown in Fig.~\ref{fig:n_a_m}. The average column density is higher than in the full filament. The fitted Plummer parameter values are slightly altered, although the results are quite close to the values obtained using the full filament.

\begin{figure*}
\includegraphics[width=16cm]{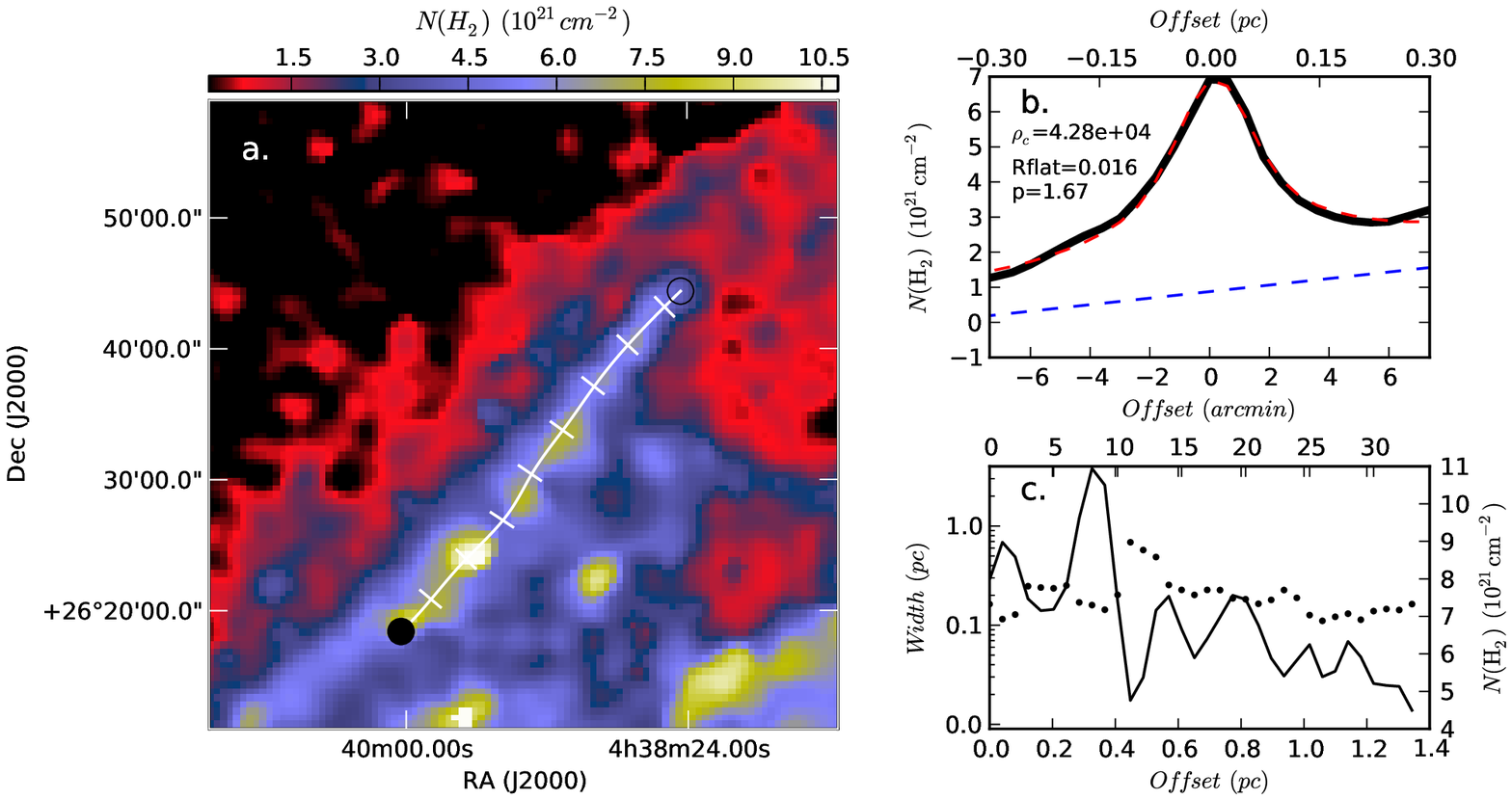}
\caption{TMC-1N column density map and filament profile derived from 2MASS extinction map. See Fig.~\ref{fig:n_a_f} for explanations of the used notation.}
\label{fig:n_a_f_2m}
\end{figure*}

For a comparison using lower resolution data, a column density map and filament profile derived from the 2MASS extinction map of TMC-1N are shown in Fig.~\ref{fig:n_a_f_2m}. Here, the filament position obtained from the \emph{Herschel} map also fits the filament seen in 2MASS extinction map very well. However, along the filament, the column density peaks are in different positions in maps derived from the \emph{Herschel} and 2MASS observations. Possible anticorrelation between FWHM and column density along the ridge of the filament is not evident using the 2MASS data. The column density derived from the 2MASS data is notably smaller than the column density derived from the \emph{Herschel} or WFCAM data. However, we also obtain a similar Plummer profile using 2MASS data.

\begin{figure*}
\includegraphics[width=16cm]{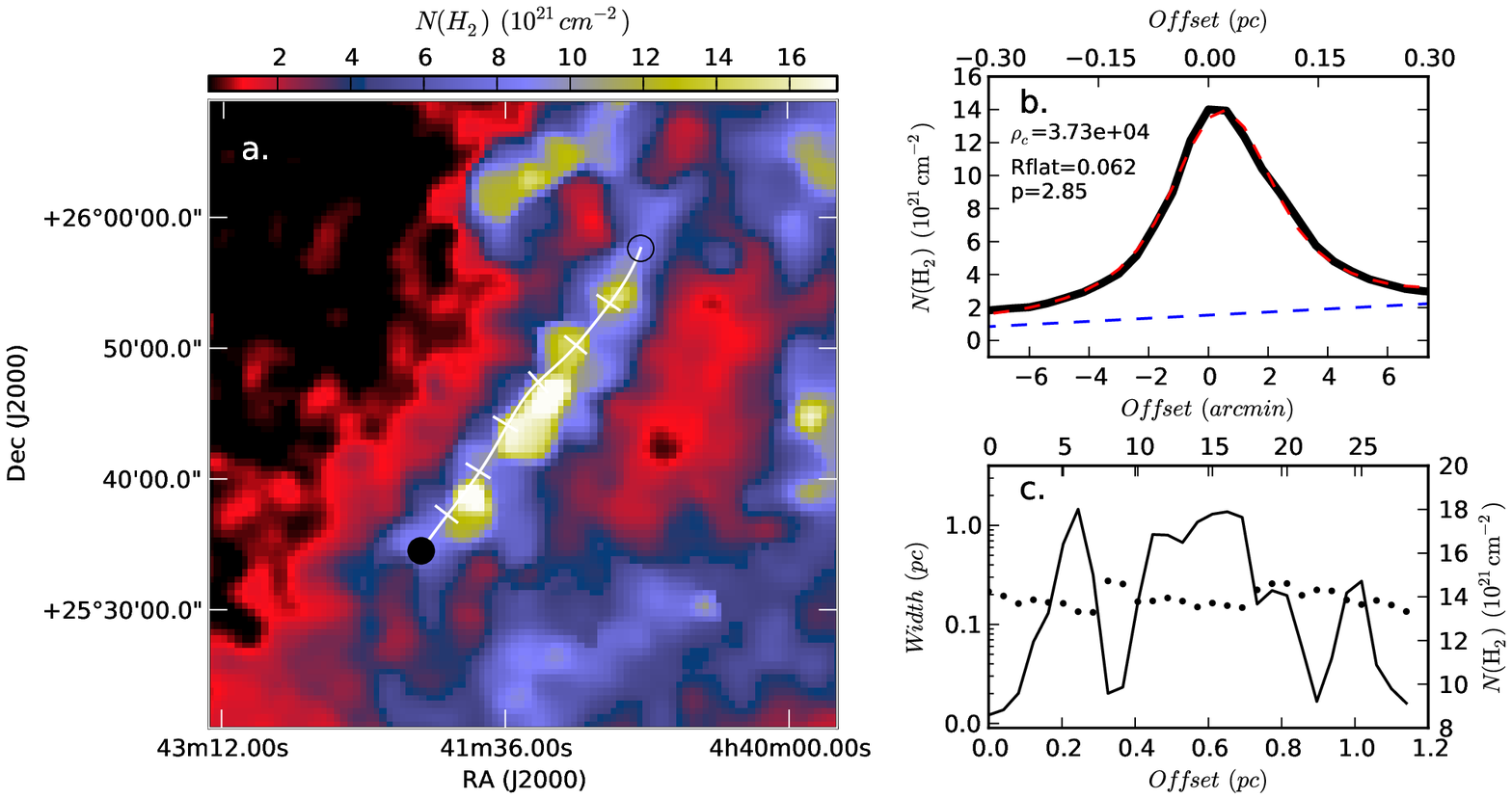}
\caption{TMC-1 column density map and filament profile derived from 2MASS extinction map (YSOs removed). See Fig.~\ref{fig:n_a_f} for explanations of the used notation.}
\label{fig:a_f}
\end{figure*}

\begin{figure*}
\includegraphics[width=16cm]{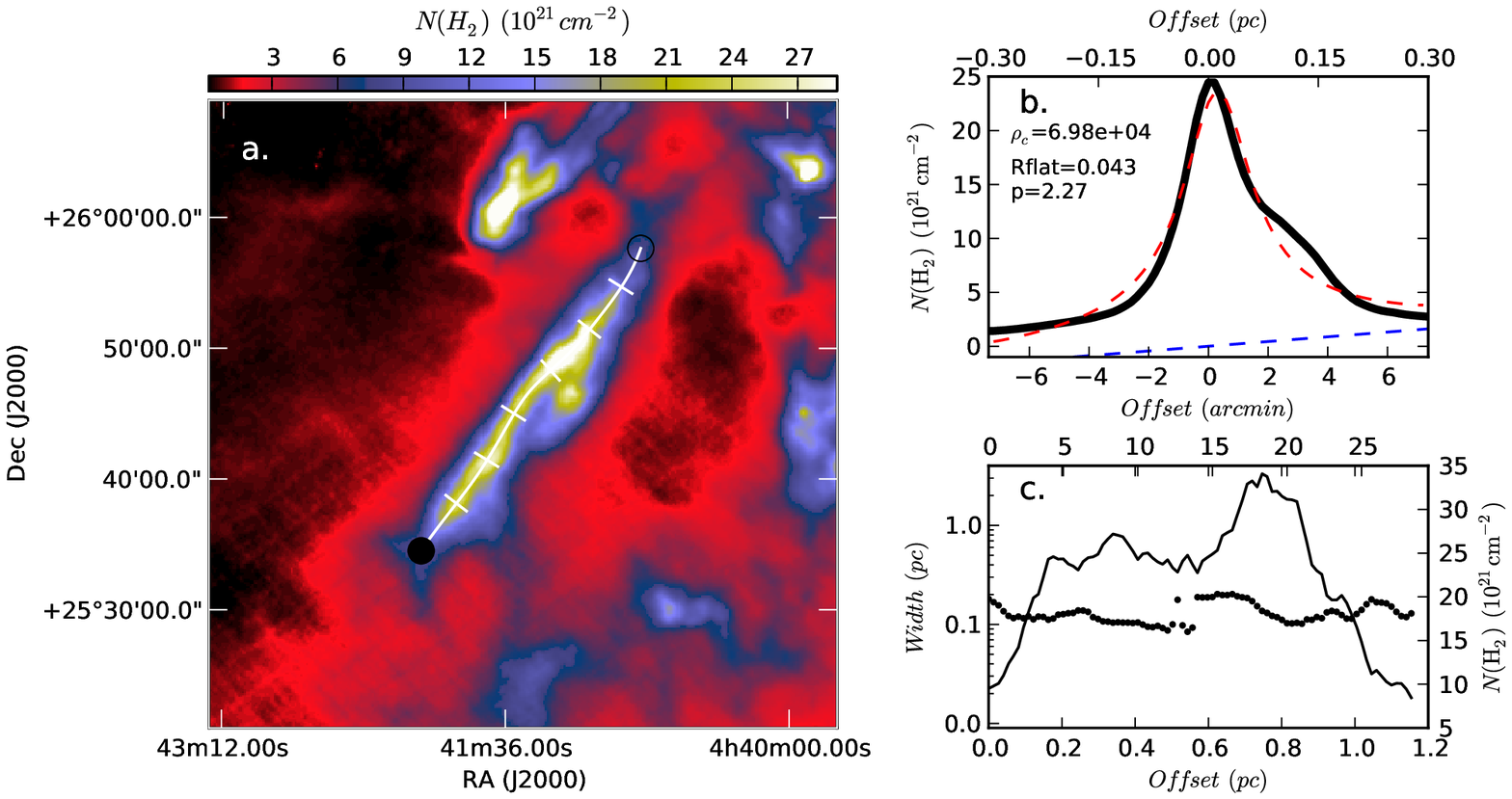}
\caption{TMC-1 column density map and filament profile derived from \emph{Herschel} dust emission map. See Fig.~\ref{fig:n_a_f} for explanations of the used notation.}
\label{fig:a_f_c}
\end{figure*}

For a comparison of another filament, column density maps and filament profiles derived from the 2MASS extinction map (YSOs removed) and \emph{Herschel} emission map of TMC-1 are shown in Figs.~\ref{fig:a_f} and \ref{fig:a_f_c}. Here the filament is again traced using the \emph{Herschel} column density map. In this case, the filament position does not exactly match the filament ridge in the 2MASS map. However, this is not a problem in the profiling process, as the filament is allowed to have a small offset from the given position. In this filament, we do not see any notable background gradient or anticorrelation between FWHM and column density along the ridge of the filament. The FWHM of the filament is approximately 0.1-0.2 pc. The column density structure along the ridge differs notably in these two data. In the \emph{Herschel} data, we see a continuous high density ridge, while in the 2MASS data there is more variation, 
because of the low number of background stars. Based on the \emph{Herschel} data, next to the filament there is a distinct clump, which effects the profile and the derived FWHM values. This change in the FWHM values is not seen in the 2MASS data.
For a comparison, we show the column density map and filament profiles derived from the 2MASS extinction map before removing YSOs in Fig.~\ref{fig:a_f_ysos}.

The obtained parameters describing the filament, namely FWHM, Plummer parameters and the mass per unit length $M_{\rm line}$ derived from observations of TMC-1N (full and middle filament) and TMC-1 (before and after removing YSOs)
using different methods are shown in Table~\ref{tab:result_param_table}.

We can compare our results for TMC-1 derived from the \emph{Herschel} maps with those of \citet{Nutter2008}, from Submillimetre Common User Bolometer Array (SCUBA)~\citep{Holland1999} and the Infrared Space Observatory ISO~\citep{Kessler1996} ISOPHOT instrument~\citep{Lemke1996} data. We note
first that we agree on the fact that the Bull's tail, as they call
it, is a filament and not a ring. Both analyses fit Plummer-like profiles,
and therefore the results should be comparable (see Table~\ref{tab:result_param_table}).
We obtain a value for $p$ of $\sim$2.3, whereas they fix $p$=3. We find
$\rho_{c} \sim 7 \times 10^4$, compared to their range of
$\rho_{c} \sim 7 \times 10^4$ to $\sim 1.7 \times 10^5$ (with
nominal values towards the upper end of this range). Our value for
$R_{flat}$ is $\sim 0.04$pc, and their value is $\sim 0.05$pc.
We find the FWHM to be 0.12pc for the \emph{Herschel} data, and $\sim$0.18pc for the 2MASS data. 
They fitted the filament with two components, the first of which had a FWHM of 0.11pc and the
second of which had a FWHM of 0.16pc. It can therefore be seen that there is a high degree of consistency, giving us added confidence in our parameter fitting of the
data.

The WFCAM $A_{\rm V}$ maps and \emph{Herschel} column density maps give very similar results for the FWHM of the TMC-1N filament (0.092 and 0.093 pc, respectively). Lower resolution $A_{\rm V}$ maps based on 2MASS stars give slightly higher values for FWHM (0.17 pc). The behaviour of the Plummer parameters or $M_{\rm line}$ derived from these is not as predictable, as there are dependencies between the parameters. We examine these dependencies more closely below.
In addition, the exact values seem to be quite sensitive for instance to small changes in the data, different resolution data, fitting algorithm, initial fitting parameter values or different profiling method. We also noticed that the chosen pixel size for the extinction maps can alter the results.

\begin{figure*}
\includegraphics[width=16cm]{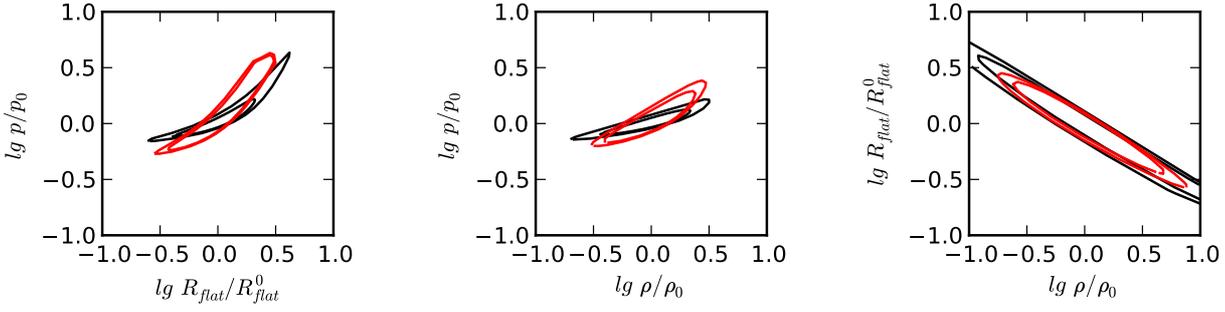}
\caption{Plummer profile parameter correlations. Contours 1.5 $\chi^2_{min}$ and 10 $\chi^2_{min}$ are drawn for Plummer profile fits obtained from WFCAM $A_{\rm V}$ map (black lines) and \emph{Herschel} column density map (red lines). In each case we fix one of the three Plummer parameters ($\rho_{\rm c}$, $R_{\rm flat}$ or $p$) and change the others.
The axes show parameter values relative to those of the $\chi^2_{min}$ (see Table~\ref{tab:result_param_table} for the precise values of the parameters).}
\label{fig:chi2_params}
\end{figure*}

In order to study the correlations of the Plummer profile parameters we save cross-sections of $\chi^2$.
We fix each of the parameters $\rho_{\rm c}$, $R_{\rm flat}$ or $p$ in turn to the fitted value, and change the others systematically to extract the $\chi^2$ values in three orthogonal parameter planes crossing at the location of the $\chi^2$ minimum, $\chi^2_{min}$.
In Fig.~\ref{fig:chi2_params}, we show the contour levels 1.5 $\chi^2_{min}$ and 10 $\chi^2_{min}$, for each case. The results show that there is a banana shaped correlation between $R_{\rm flat}$ and $p$, more or less linear (in log-space) correlation between $\rho_{\rm c}$ and $p$, and a linear anticorrelation between $\rho_{\rm c}$ and $R_{\rm flat}$. 
Even the 1.5 $\chi^2_{min}$ contours are quite wide, especially in $\rho_{\rm c}$--$R_{\rm flat}$-plane,
meaning that very different parameter values can give small $\chi^2$ values. 
There are also differences in the size and orientation of the contours when comparing \emph{Herschel} emission maps and WFCAM extinction maps. The 1.5 $\chi^2_{min}$ contours are smaller when using WFCAM extinction map in 
$R_{\rm flat}$--$p$-plane and $\rho_{\rm c}$--$p$-plane.
However, in $\rho_{\rm c}$--$R_{\rm flat}$-plane
the contours are smaller when using \emph{Herschel} emission maps.

\begin{table*}
\caption{Obtained parameters describing the filaments TMC-1N (full and middle filament) and TMC-1. The columns are the data set, method (column density map ($A$), $A_{\rm V}$ map ($B$), median profile ($C$) and fit to individual stars ($D$)),
FWHM (including the effect of the data resolution), Plummer parameters and mass per unit length $M_{\rm line}$.}
\label{tab:result_param_table}
\begin{tabular}{lllllll}
\hline \hline
Data set & Method & FWHM & $\rho_{\rm c}$ & $R_{\rm flat}$ & $p$ & $M_{\rm line}$\\
~                 &~       & (pc) & (cm$^{-3}$)    & (pc)           &~    & ($M_{\sun}$/pc)\\
\hline
TMC-1N: full filament & & & & & & \\   
\hline
\emph{Herschel} maps & $A$ & 0.093 & 3.94e+04 & 0.035 & 2.81 & 1.91e+01\\
WFCAM: $A_{\rm V}$ maps & $B$ & 0.092 & 9.02e+04 & 0.012 & 1.84 & 3.04e+01\\
2MASS: $A_{\rm V}$ maps & $B$ & 0.17 & 4.28e+04 & 0.016 & 1.67 & 3.40e+01\\ 
WFCAM: Median profile & $C$ & 0.12 & 1.70e+04 & 0.074 &  4.57 & 1.30e+01\\
WFCAM: Fit to individual stars & $D$ & ... & 2.17e+04 &  0.058 & 3.71 & 1.49e+01\\
2MASS: Median profile & $C$ & 0.16 & 1.55e+04 & 0.048 &  2.44 & 1.98e+01\\
2MASS: Fit to individual stars & $D$ & ... & 1.62e+04 &  0.068 & 3.76 & 1.49e+01\\
\hline
TMC-1N: middle filament & & & & & & \\
\hline
\emph{Herschel} maps & $A$ & 0.088 & 5.49e+04 & 0.028 & 2.37 & 2.93e+01\\
WFCAM: $A_{\rm V}$ maps & $B$ & 0.088 & 1.67e+05 & 0.007 & 1.56 & 5.95e+01\\
2MASS: $A_{\rm V}$ maps & $B$ & 0.21 & 1.78e+04 & 0.044 & 1.93 & 3.97e+01\\ 
WFCAM: Median profile & $C$ & 0.15 & 1.70e+04 & 0.063 & 2.92 & 2.33e+01\\
WFCAM: Fit to individual stars & $D$ & ... & 4.33e+04 & 0.023 & 1.79 & 4.40e+01\\
2MASS: Median profile & $C$ & 0.18 & 1.35e+04 & 0.087 & 3.38 & 2.49e+01\\
2MASS: Fit to individual stars & $D$ & ... & 2.66e+04 &  0.028 & 1.63 & 5.55e+01\\
\hline
TMC-1 & & & & & & \\
\hline
\emph{Herschel} maps & $A$ & 0.12 & 6.98e+04 & 0.043 & 2.27 & 9.02e+01\\
2MASS: $A_{\rm V}$ maps (with YSOs) & $B$ & 0.17 & 3.91e+04 & 0.078 & 3.60 & 5.06e+01\\
2MASS: $A_{\rm V}$ maps (without YSOs) & $B$ & 0.18 & 3.73e+04 & 0.062 & 2.85 & 5.24e+01\\
2MASS: Median profile & $C$ & 0.05 & 9.60e+04 & 0.019 & 1.84 & 6.56e+01\\
2MASS: Fit to individual stars & $D$ & ... & 6.56e+04 & 0.032 & 1.95 & 8.23e+01\\
\hline
\end{tabular}
\end{table*}

\subsection{Profiles using $A_{\rm V}$ of stars directly}

\begin{figure}
\includegraphics[width=9cm]{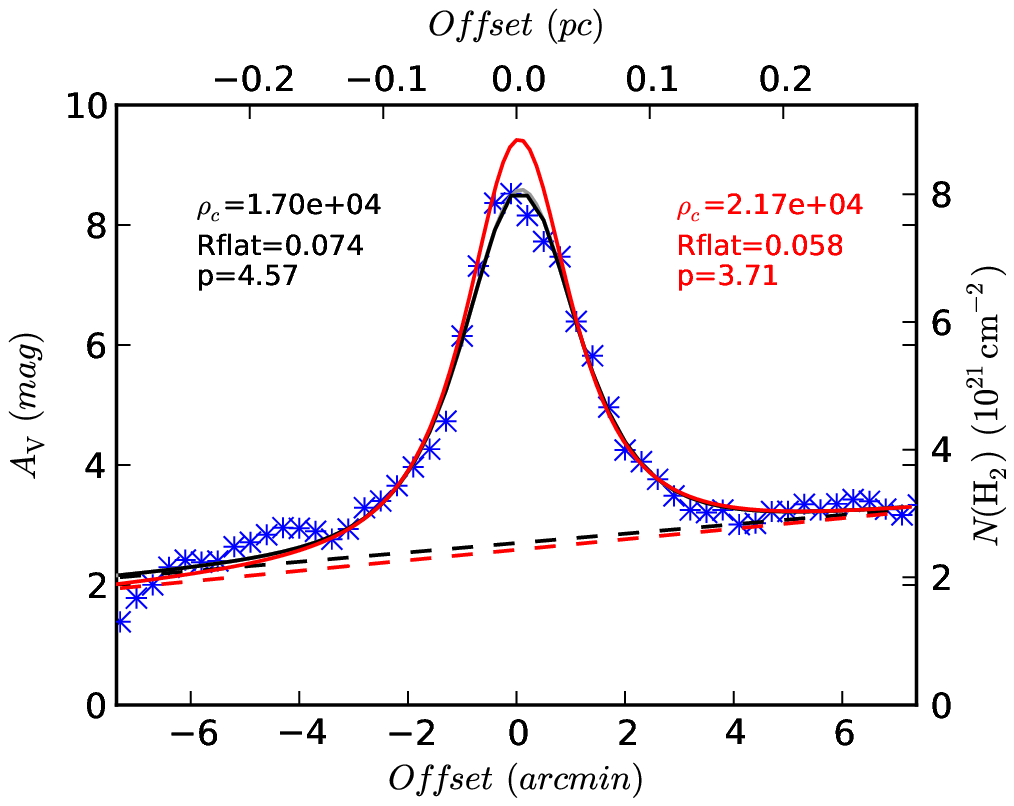}
\includegraphics[width=9cm]{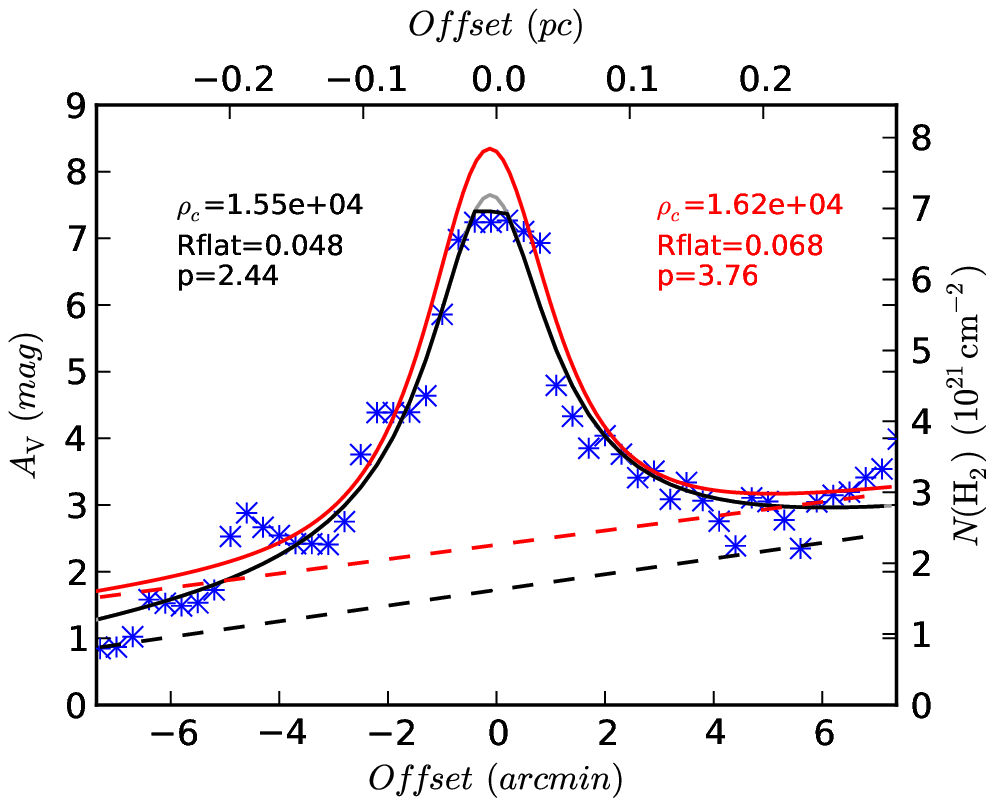}
\caption{Profiles of TMC-1N filament derived by using methods $C$ and $D$ in WFCAM data (top frame) and 2MASS data (bottom frame). Blue stars: median values. Gray line: Plummer profile before convolution, fitted to the median $A_{\rm V}$ values of stars. Black solid line: The convolved Plummer profile. Red solid line: fit to individual stars (not marked in the picture). Dashed lines: fitted linear baselines for Plummer fits shown with the same colours.}
\label{fig:stars}
\end{figure}

\begin{figure}
\includegraphics[width=9cm]{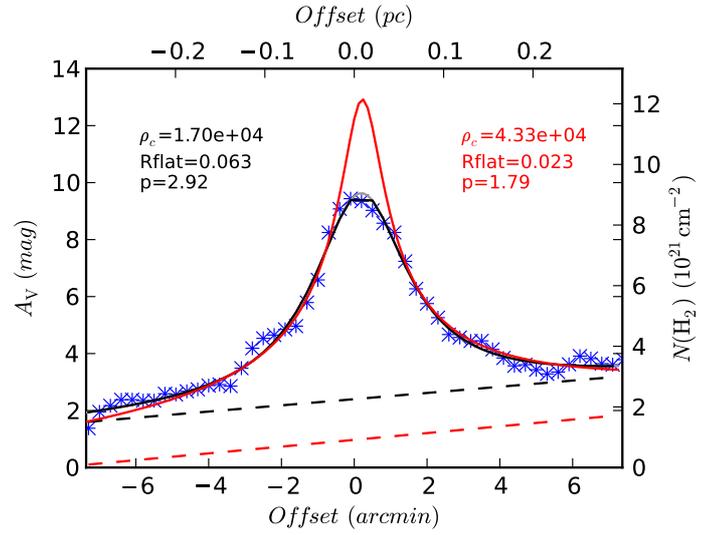}
\caption{Profile of TMC-1N middle filament derived by using individual stars or their median averages in WFCAM data. See Fig.~\ref{fig:stars} for explanations of the used notation.}
\label{fig:stars_N_middle}
\end{figure}

\begin{figure}
\includegraphics[width=9cm]{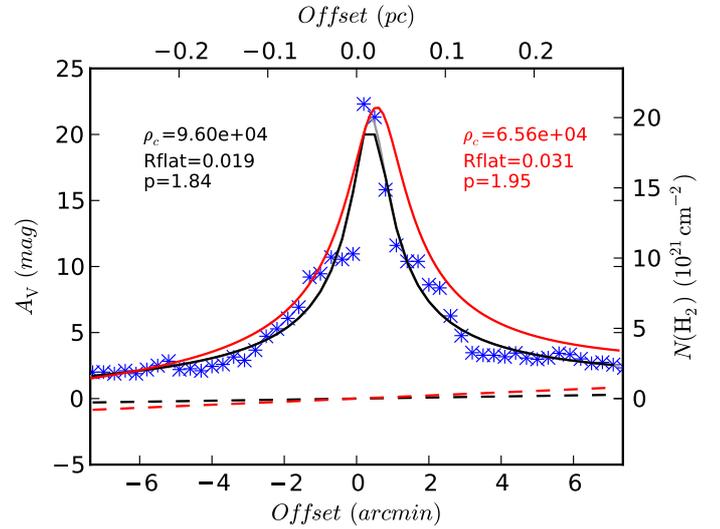}
\caption{Profiles of TMC-1 filament derived by using individual stars or their median averages in 2MASS data. See Fig.~\ref{fig:stars} for explanations of the used notation.}
\label{fig:stars_TMC1}
\end{figure}

Profiles of TMC-1N full filament derived by using methods $C$ and $D$ with WFCAM data and 2MASS data are shown in Fig.~\ref{fig:stars}. The profile of TMC-1N middle filament using the WFCAM data is shown in Fig.~\ref{fig:stars_N_middle}, and the profile of TMC-1 using the 2MASS data is shown in Fig.~\ref{fig:stars_TMC1}.
The obtained parameters derived from observations of TMC-1N (full and middle filament, WFCAM and 2MASS data) and TMC-1 (2MASS data) are shown in Table~\ref{tab:result_param_table}. The results of using median averages (method $C$) are on the 'Median profile' and the results of using individual stars (method $D$) on the 'Fit to individual stars' line. 
The results obtained for TMC-1 before and after removing the YSOs were identical when using methods $C$ and $D$.

The values obtained with methods $C$ and $D$ are quite similar to the values derived from maps. However, as with the map method, the derived parameters are quite sensitive to small changes in the data. For example, for the TMC-1N full filament using the WFCAM data, the star methods give notably larger values for $p$ than the map methods, but when only the densest middle part of the filament is profiled, these differences get smaller.

\section{Simulated extinction observations} \label{sect:simulations}

We use simulations to examine the accuracy with which the filament
profiles can be estimated from the observations of background stars.
As a reference, we use the WFCAM observations from which we estimate
the probability distribution of the number of stars as a function of
the H-band magnitude, the average J-H and H-K colours, and the
uncertainties of these colours as a function of the magnitude. These
data are used in simulations in which we place background stars at random
positions over an area of 30$\arcmin\times$30$\arcmin$. As a
reference, we use the stellar density in the WFCAM observations but
also test the effect of lower and higher stellar densities. The
stellar colours are reddened according to the column density
distribution that results from the presence of a filament. The density
distribution of the filament follows the Plummer function with chosen
parameters. The filament is assumed to be in the plane of the sky but
has a position angle of 30$^\circ$ in order to avoid the special case
where the filament would be aligned with the axes used in the
pixelization of the extinction maps.

The extinctions of the individual stars (and of the
30$\arcmin\times$30$\arcmin$ extinction map) are calculated using one
corner (78 square arc minutes) of the simulated region as a reference
area from which the intrinsic stellar colours are estimated. The
simulated observations are then analysed as in the case of the WFCAM
observations above. The filament parameters are derived from the
extinction map (method $B$), from the median profile along the filament (method $C$), and from
the direct fit of the Plummer profile to the $A_{\rm V}$ values of the
individual stars (method $D$). We analyse a 20$\arcmin$ long part of the filament
that thus fits well within the 30$\arcmin\times$30$\arcmin$ area with
the assumed distance of 140\,pc, even taking into account the non-zero
position angle.

In the simulations, $R_{\rm flat}$ is always 0.025, and there is no background gradient. We compare our three methods and the way the fitted $p$, $\rho_{\rm c}$ and derived $M_{\rm line}$ change when we use different sets of parameter values for $\rho_{\rm c}$ ($10^4$, $10^5$, $10^6$ cm$^{-3}$), $p$ (2, 3, 4), and stellar density (0.2, 0.5, 1, 2, 5, 10, where 1 denotes the stellar density in the WFCAM observations). We also test how photometric noise affects the results by using different noise levels 0, 0.3, and 1, where 1 denotes the photometric noise level in the WFCAM observations of TMC-1N. Even when the photometric noise is zero, there is still noise in the data, because the stars have a random position and brightness, i.e., some stars may be extincted below the detection threshold. 
The filament looks different depending on the number and brightness of the stars situated in the filament area. The filament is well defined, if there are enough bright stars near the borders of the filament, where they can still be seen. We run each simulation case 100 times for the estimation of the uncertainties.

\subsection{Simulation results} \label{sect:simu_results}

\begin{figure*}
\includegraphics[width=16cm]{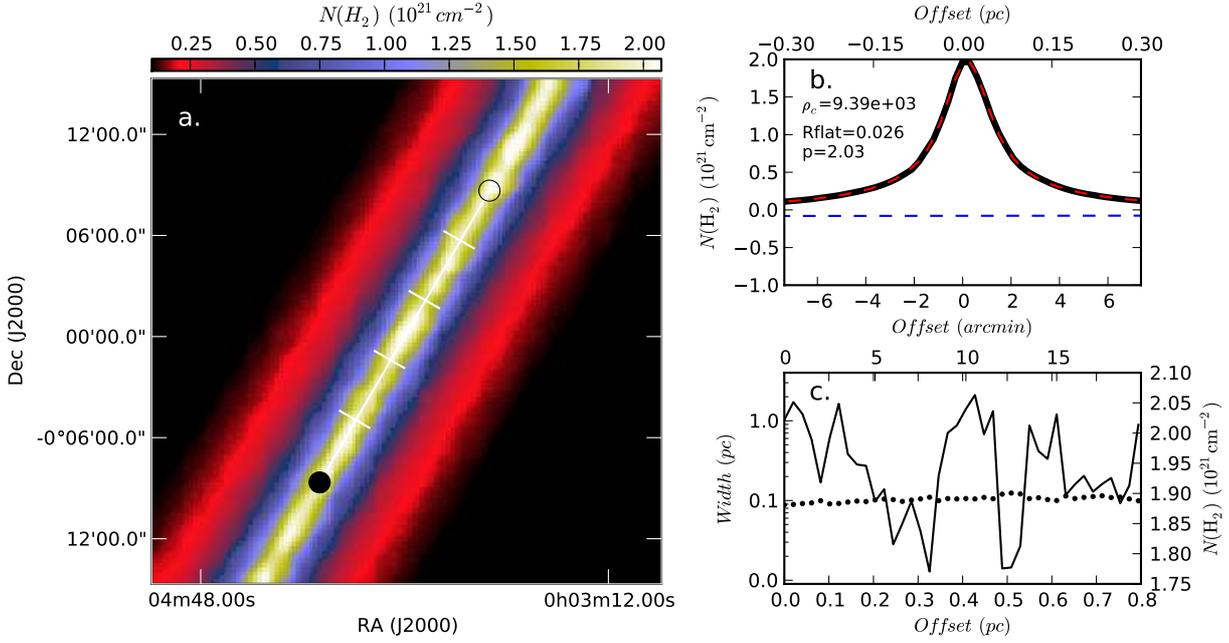}
\caption{Example of a simulated map and filament profile without noise. Used parameter values in the simulation: $\rho_{\rm c}$=$10^4$cm$^{-3}$, $R_{\rm flat}$=0.025, $p$=2.0, stellar density = 1.0, and without photometric noise. The fitted parameter values are shown in frame b. See Fig.~\ref{fig:n_a_f} for explanations of the used notation.}
\label{fig:simumap_e4}
\end{figure*}

\begin{figure*}
\includegraphics[width=16cm]{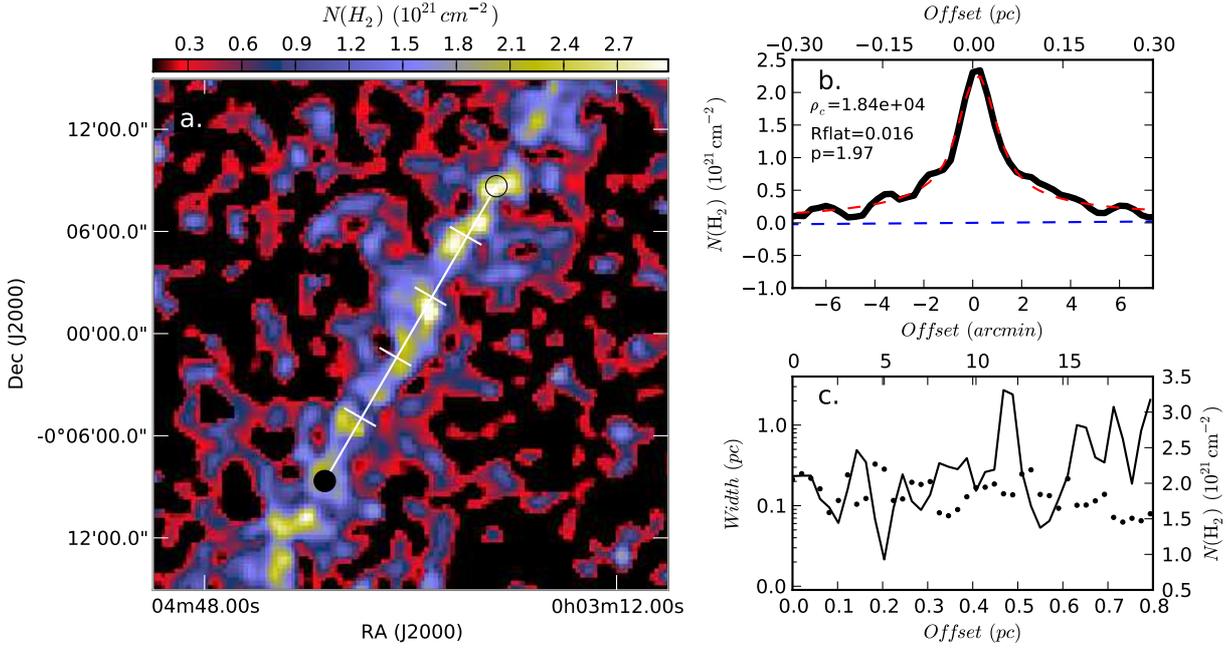}
\caption{Example of a simulated map and filament profile with noise. Used parameter values: $\rho_{\rm c}$=$10^4$cm$^{-3}$, $R_{\rm flat}$=0.025, $p$=2.0, stellar density = 1.0, and noise=1.0 (meaning a similar noise level as in our WFCAM observations of TMC-1N.). The fitted parameter values are shown in frame b. See Fig.~\ref{fig:n_a_f} for explanations of the used notation.}
\label{fig:simumap_e4_n1}
\end{figure*}

\begin{figure*}
\includegraphics[width=8cm]{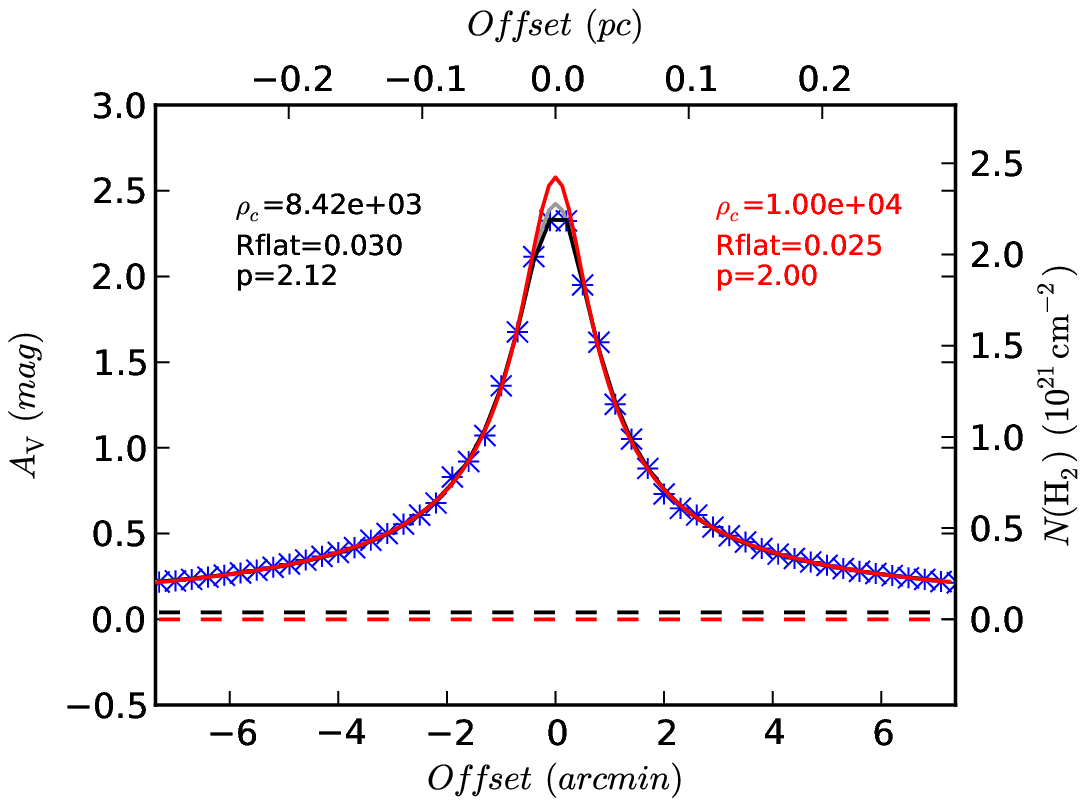}
\includegraphics[width=8cm]{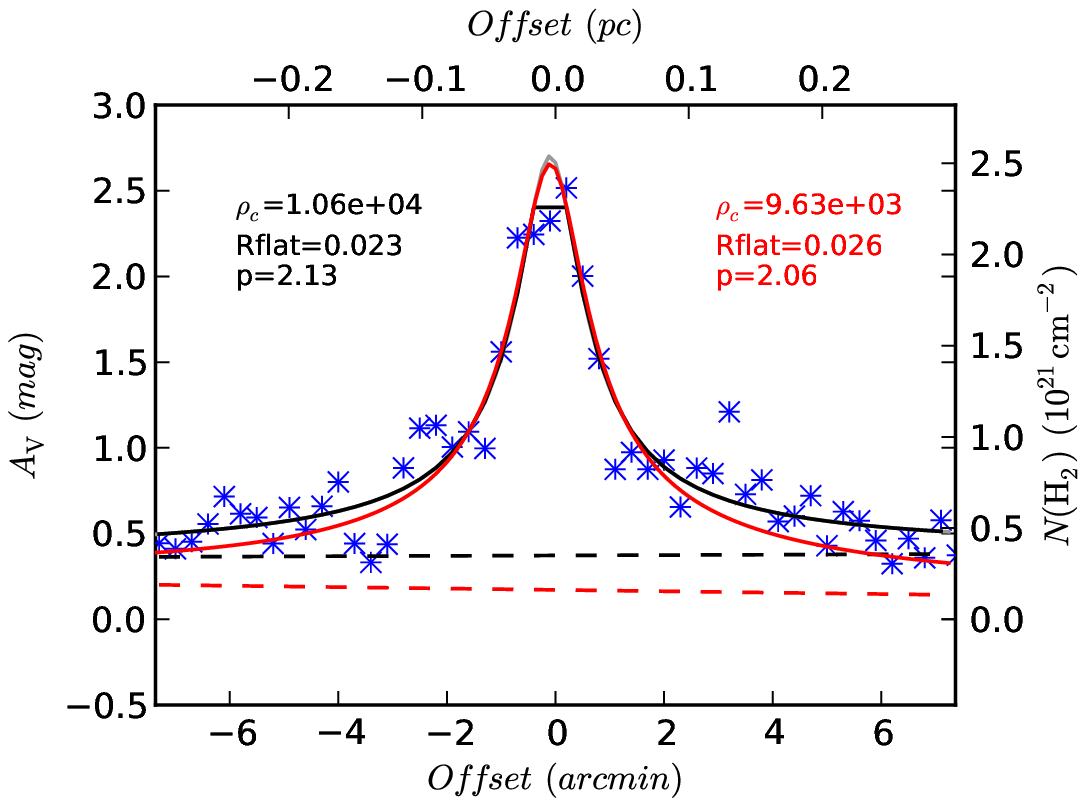}
\caption{Examples of simulated profiles using methods $C$ and $D$. Used parameter values: $\rho_{\rm c}$=$10^4$cm$^{-3}$, $R_{\rm flat}$=0.025, $p$=2.0, stellar density = 1.0, and without noise (left frame) or with noise (right frame). The fitted parameter values are shown in the figure. See Fig.~\ref{fig:stars} for explanations of the used notation.}
\label{fig:simustar_e4}
\end{figure*}

Examples of simulated maps and filament profiles with and without photometric noise are shown in Figs.~\ref{fig:simumap_e4} and \ref{fig:simumap_e4_n1}. The simulated profiles of the same cases using methods $C$ and $D$ are shown in Fig.~\ref{fig:simustar_e4}. The resulting column density maps resemble our TMC-1N observations, except that the density distribution precisely follows the Plummer function with a chosen set of parameters. If all the simulated filaments are examined along their whole length, there is no clear anti-correlation (nor correlation) between observed FWHM and column density along the ridge. However, similarly to the observations of TMC-1N (Fig.~\ref{fig:n_a_f}, frame c), anticorrelation can sometimes be seen in some of the densest peaks also in simulations, even though the simulated filament itself is symmetric.

The filament parameters are derived from the simulated observations. 
In Figs. 15--20,
we compare the true and estimated parameter values $p$, $\rho_{\rm c}$, and  $M_{\rm line}$ for cases with different parameter values and different number of background stars. The error bars correspond to half of the interquartile range between the 75\% and 25\% percentiles.

\begin{figure*}
\centering
\includegraphics[width=13cm]{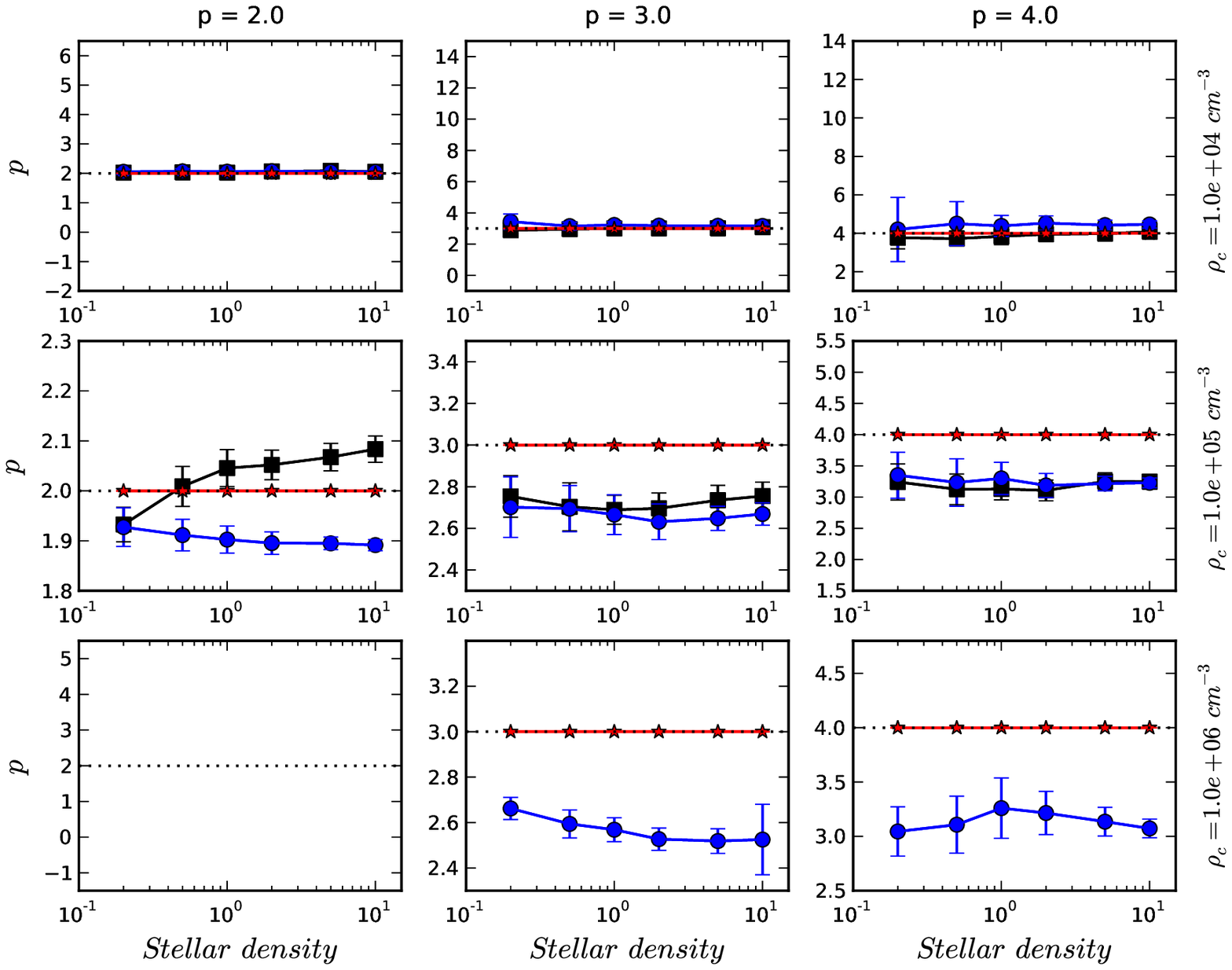}
\caption{Plummer $p$ parameter derived from simulated extinction observations without noise. Different central densities are shown on different lines: $10^4$ (cm$^{-3}$) (uppermost frames), $10^5$ (middle frames), and $10^6$ (bottom frames). Symbols and colours denote the used method: method $B$ (\textit{black square}), method $C$ (\textit{blue circle}), and method $D$ (\textit{red star}). Stellar density is given on the x-axis, where 1 means a similar density as in the WFCAM observations of TMC-1N. The true $p$ value (2, 3 or 4) used in the simulations is given on top of each column and also shown as a dashed line in the figures. The derived $p$ value is marked in the figures for each simulated stellar density and method with scale on the y-axis.}
\label{fig:p0}
\end{figure*}

Figure~\ref{fig:p0} shows that in a noiseless situation method $D$ works almost perfectly in fitting the $p$ parameter. The error bars of method $D$ are not visible in the figure, because they are so small.
This is natural as the random positions of the stars do not affect this method and the photometry itself does not contain errors.
Methods $B$ and $C$ also work well for modest density, $\rho_{\rm c}$ = $10^4$ cm$^{-3}$, being correct to within 10\%.
However, in higher density ($\rho_{\rm c}$ = $10^5$ cm$^{-3}$) both of these methods underestimate the $p$ value, except in the case $p$ = 2, where method $B$ ends up overestimating the $p$ value. Still, both methods are correct to within 20\%.
Method $B$ does not work reliably at densities as high as $10^6$ cm$^{-3}$, because stars cannot be seen at the centre of the filament. It is therefore left out of the figures.
The case of density $10^6$ cm$^{-3}$ and $p$ = 2 is not shown either with methods $C$ and $D$, as the results get unreliable when the empty part in the middle is too wide.
However, with larger $p$ values both methods $C$ and $D$ continue to work also at high densities, although method $C$ underestimates the $p$ value $\sim$20\%.
We also made some tests with density $10^3$ cm$^{-3}$, but for such low density the filament is no longer reliably identified. The combination of central density and true parameter $p$ determines how well the observed parameter $p$ is fitted.

\begin{figure*}
\centering
\includegraphics[width=13cm]{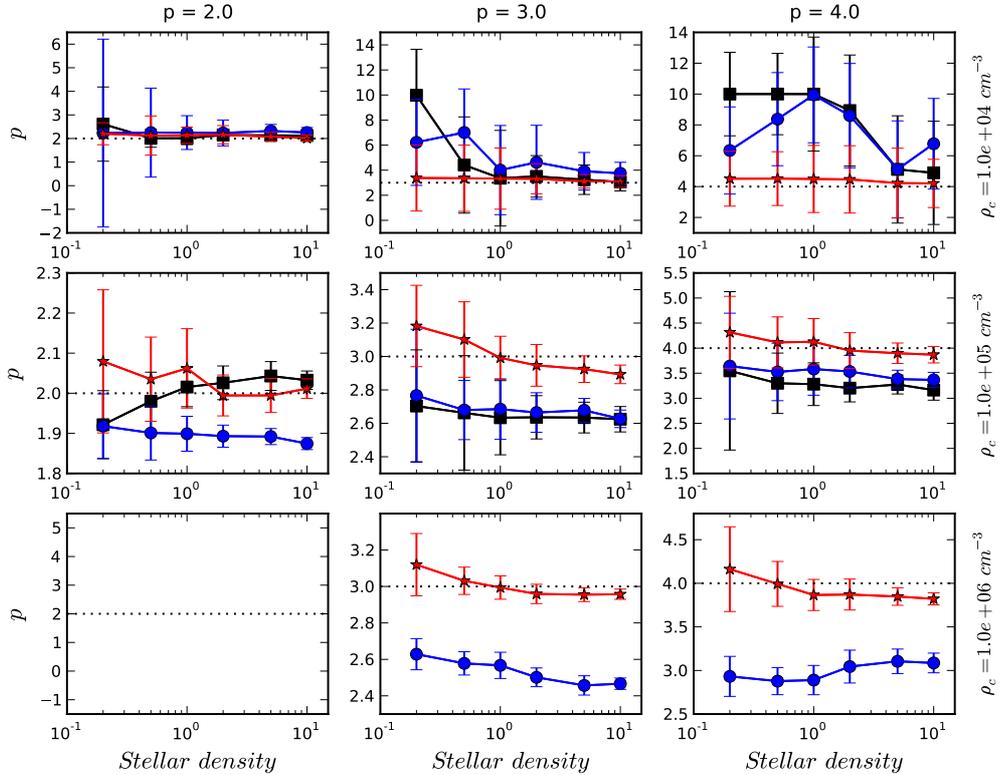}
\caption{Plummer $p$ parameter derived from simulated extinction observations with noise = 1 (times WFCAM noise). See Fig.~\ref{fig:p0} for explanations of the used notation.}
\label{fig:p1}
\end{figure*}

The results with photometric noise similar to WFCAM observations are shown in Fig.~\ref{fig:p1}. The results are still similar to those without noise, except that the errors get larger. For density 10$^4$\,cm$^{-3}$ and parameter $p$=2.0 the bias is very small with all star densities and all methods, but in the case of very low star density (0.2), the scatter for method $C$ is so large that the method does not work reliably.
For density 10$^4$\,cm$^{-3}$ and parameter $p$=4.0, both methods $B$ and $C$ typically result in a value of $p=10$, the maximum allowed in the fitting procedure. This means that in these cases the results of the fits are not reliable.
Still, in most cases the results are correct to within 10\% for method $D$ and to within 20\% for the other methods.

\begin{figure*}
\centering
\includegraphics[width=14cm]{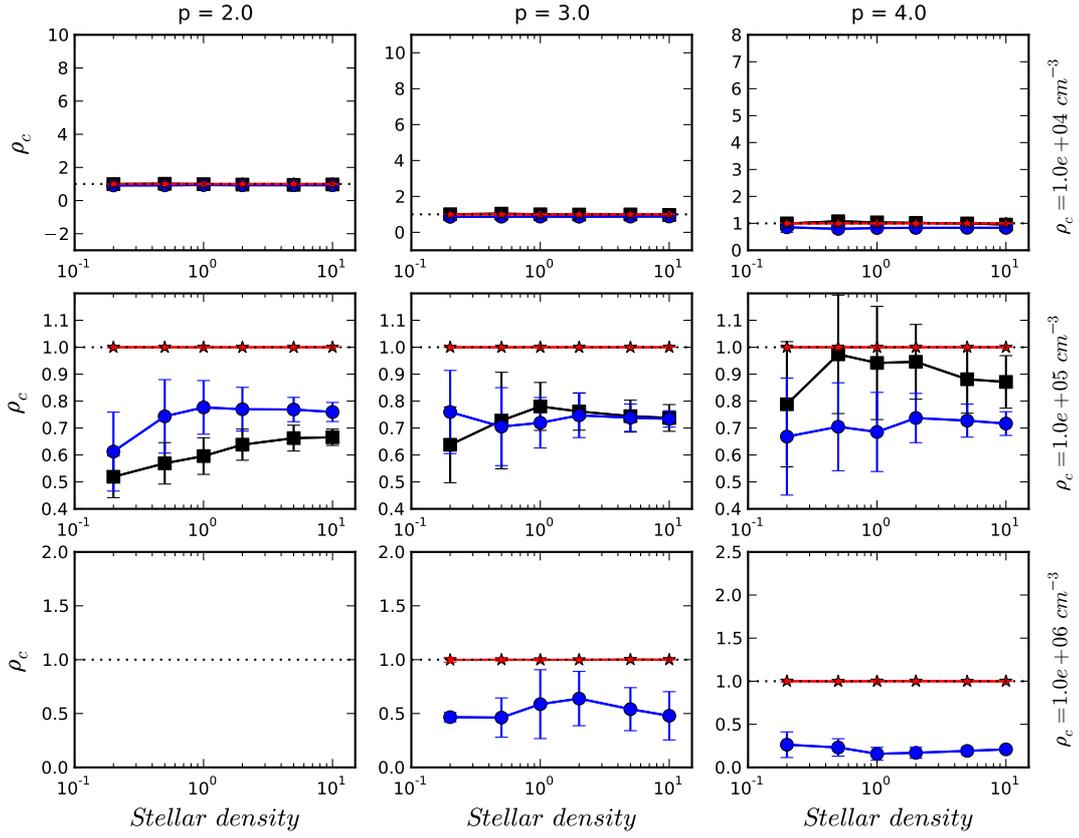}
\caption{Plummer $\rho_{\rm c}$ parameter derived from simulated extinction observations without noise. Y-axis denoting the fitted $\rho_{\rm c}$ is divided with the correct $\rho_{\rm c}$. See Fig.~\ref{fig:p0} for explanations of the used notation.}
\label{fig:rho0}
\end{figure*}

\begin{figure*}
\centering
\includegraphics[width=14cm]{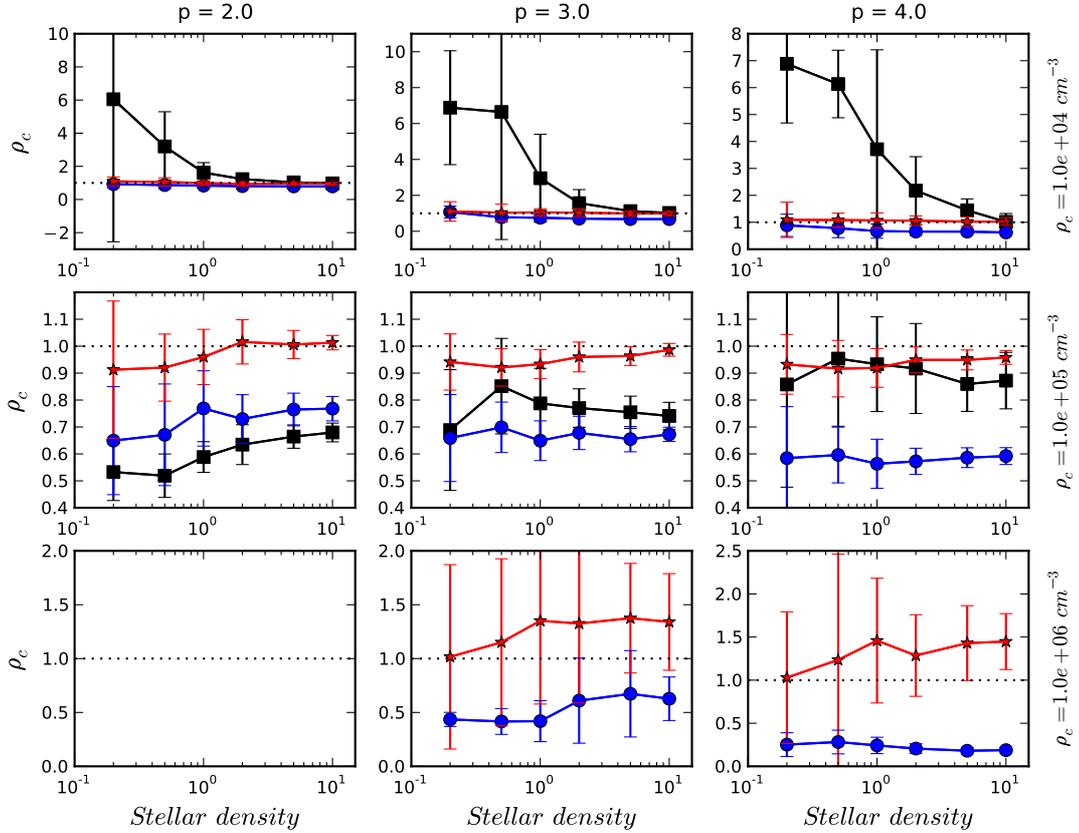}
\caption{Plummer $\rho_{\rm c}$ parameter derived from simulated extinction observations with noise = 1. See Figs.~\ref{fig:p0} and \ref{fig:rho0} for explanations of the used notation.}
\label{fig:rho1}
\end{figure*}

\begin{figure*}
\centering
\includegraphics[width=14cm]{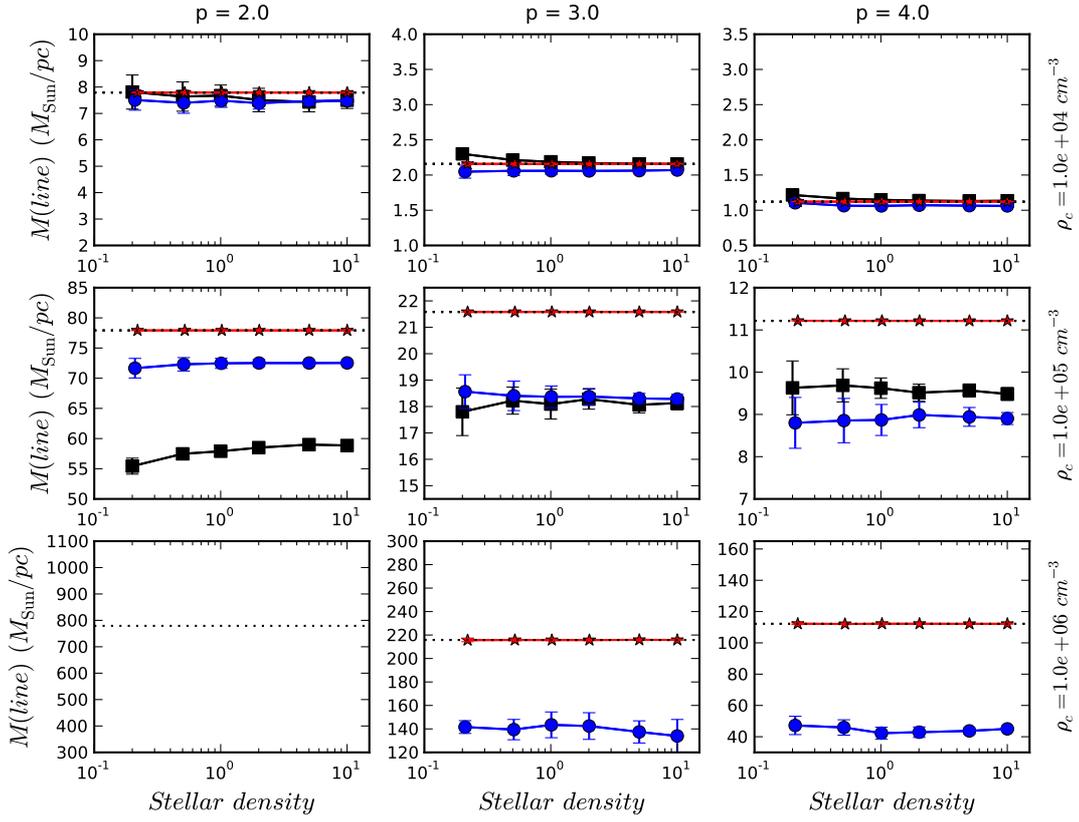}
\caption{$M_{\rm line}$ derived from simulated extinction observations with noise = 0. The true $M_{\rm line}$ is marked with a dashed line in the figures. The derived $M_{\rm line}$ is marked in the figures for each simulated stellar density and method with scale on the y-axis. Otherwise, see Fig.~\ref{fig:p0} for explanations of the used notation.}
\label{fig:m0}
\end{figure*}

\begin{figure*}
\centering
\includegraphics[width=14cm]{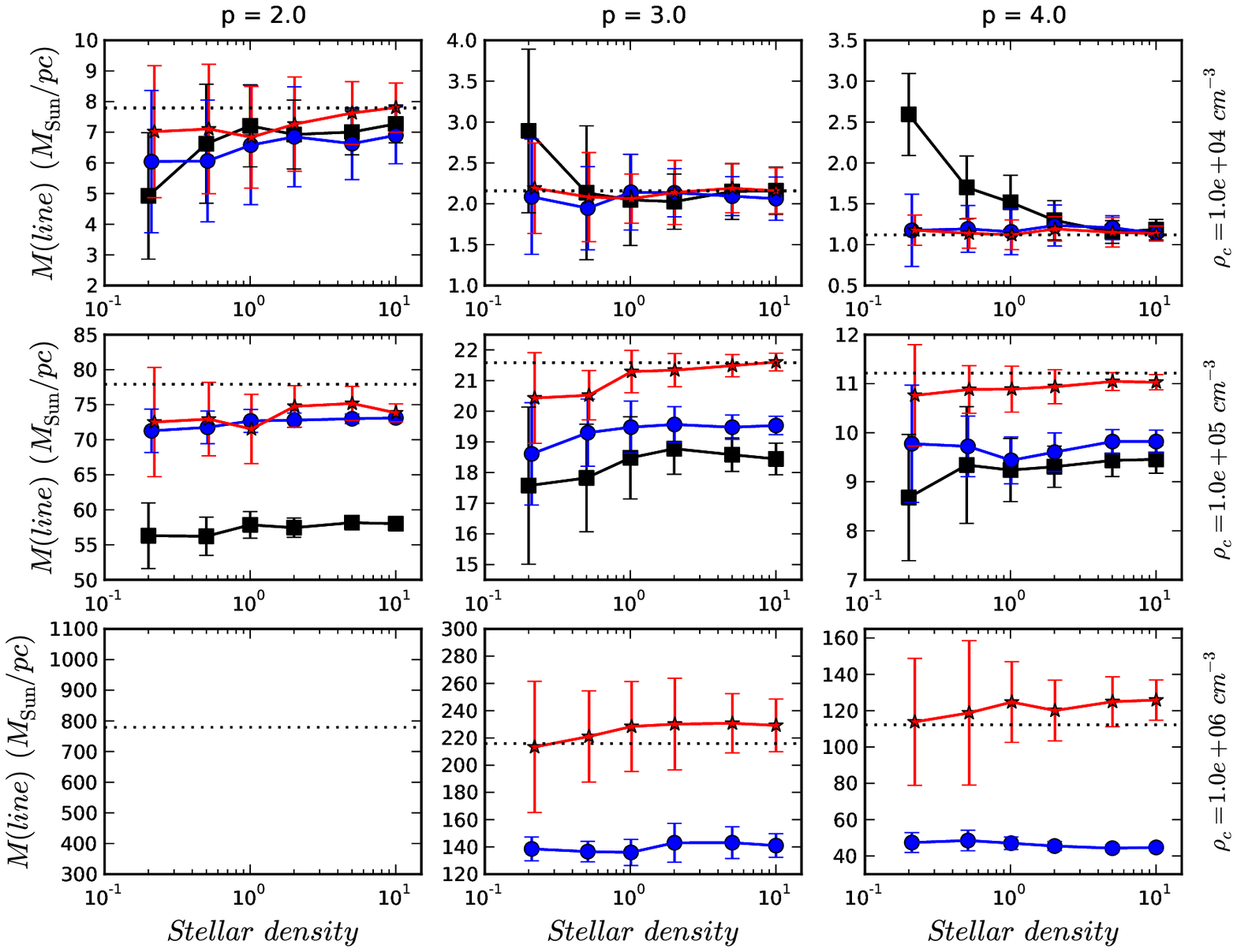}
\caption{$M_{\rm line}$ derived from simulated extinction observations with noise = 1. See Figs.~\ref{fig:p0} and \ref{fig:m0} for explanations of the used notation.}
\label{fig:m1}
\end{figure*}

Figures~\ref{fig:rho0} and \ref{fig:rho1} show, that $\rho_{\rm c}$ is also fitted quite reliably, especially using methods $D$ and $C$ for modest ($\rho_{\rm c} \sim$ $10^4$ cm$^{-3}$) densities and without photometric noise. If noise is included, method $B$ tends to overestimate the density for the modest density case, especially when the star density is low. For higher densities ($\rho_{\rm c} >=$ $10^5$ cm$^{-3}$), both methods $B$ and $C$ underestimate the density, even though the results are still correct to within an order of magnitude. The derived $M_{\rm line}$ shows a similar behaviour in Figs.~\ref{fig:m0} and ~\ref{fig:m1}. Method $D$ gives correct results to within 10\% in most cases. Methods $B$ and $C$ also give correct results to within 10\% for modest density ($\rho_{\rm c} \sim$ $10^4$ cm$^{-3}$).
For higher densities, methods $B$ and $C$ both underestimate the mass by $\sim$20\% or even up to $\sim$60\% in the densest case.

Generally, all the compared methods seem to produce reliable results in most cases, especially with modest density ($10^4$ cm$^{-3}$). In addition, when stellar density grows, the errors become smaller.
For higher densities, method $D$ usually seems to give the most reliable value.

\clearpage

\section{Discussion} \label{sect:discussion}

The characteristics of interstellar filament properties are currently approached using mainly submillimetre studies, due to the high resolution data obtained with \emph{Herschel}. However, if such data are not available, the use of groundbased NIR data would be more convenient. We have therefore compared the profiles obtained from column density maps derived from both submillimetre dust emission maps (method $A$) and NIR extinction maps (method $B$).

We have presented two new methods to derive filament profiles. In method $C$ we divide the stars to bins according to their distance from the filament centre, calculate the median $A_V$ for each offset bin and fit a Plummer profile to the median $A_V$ values. In method $D$ we fit the Plummer profiles directly to the $A_V$ values of individual stars, without any spatial averaging.

The results of the profile fits to TMC-1N and TMC-1 (Table~\ref{tab:result_param_table}) show, that all four methods  give rather similar results. Of course, with observations we do not know which values are the most correct ones. There does not seem to be any obvious systematic biases in different methods, e.g., $A_{\rm V}$ maps can sometimes give a lower (in TMC-1) and sometimes higher (in TMC-1N) values for masses than the \emph{Herschel} emission maps. In general, the fitted densities are similar to within an order of magnitude. The fitted $p$ values change relatively much, between 1.67--4.57 for TMC-1N (full filament), 1.56--3.38 for TMC-1N (middle filament), and 1.84-3.60 for TMC-1, depending of the method.
However, in the other two filaments shown in Appendix~\ref{sect:appendix_A}, the $p$ values are much more stable (1.92--2.1 and 1.31--1.79).

YSOs embedded in the filaments could bias the extinction estimates, as they are redder than background stars. Even the effect of a single YSO could be notable in dense filaments, if only a few stars are seen in the middle of the filament.
There are six known YSOs in the TMC-1 filament area used in our profiles. We have tested the effect of YSOs in this filament, and note that removing the YSOs can have significant local effects in the column density maps and profiles along the ridge of the filament. However, in this case, the profile parameters derived with or without YSOs stay identical when using methods $C$ or $D$. Using method $B$, there is some change in the fits, but the parameter values are still similar to within $\sim$10-30\%.

We have noticed that the Plummer profile fits are quite sensitive to small changes in the data or the fitting routines. The dependencies of the Plummer parameters in Fig.~\ref{fig:chi2_params} show that the confidence regions of Plummer parameters are quite wide. This confirms our findings, that the parameters are not always well constrained and therefore small changes in the data can lead to different parameter values. The correlations of the parameters, e.g. $R_{\rm flat}$ and $p$, should therefore be taken into account, instead of giving too much weight on the value of a single parameter.
The emission and extinction data are independent and together should result in more accurate estimates of the filament parameters. This is particularly true when the orientations of the corresponding confidence regions (i.e., the minimum $\chi^2$ contours) are different.

We have used simulations to compare the filament parameters and the derived mass per unit length $M_{\rm line}$ obtained with extinction maps and the two alternative methods. We have tested the limit of the methods using different densities
$\rho_{\rm c}$ ($10^4$, $10^5$, $10^6$ cm$^{-3}$) and stellar densities (0.2, 0.5, 1, 2, 5, 10, where 1 denotes the stellar density in the WFCAM observations). Tests with $\rho_{\rm c}$ = $10^3$ showed, that in these low densities the extinction methods did not work reliably.
The simulations show that method $D$ gives correct results to within 10\% in most cases. Methods $B$ and $C$ also work in most cases correctly to within $\sim$20\%. Only for high densities (10$^6$\,cm$^{-3}$) can the bias for method $C$ be notably larger, and method $B$ cannot reliably be used. When using method $D$ in high density the scatter can grow so much that the results are unreliable, even though the bias can still be small.
When stellar density grows, the results become more accurate. However, the simulations showed that all the methods also work in most cases with stellar densities below that of our WFCAM observations.

In nearby, high galactic latitude clouds, such as Taurus, foreground stars do not cause a problem in the analysis. In other cases, however, foreground stars could cause a notable bias. We have examined the effect of foreground stars in simulations. Changing $\sim$10\% of the background stars to foreground stars can cause $\sim$10-30\% more bias in the mass estimates. The method $D$ is quite robust when no noise is included, but the difference to the other methods is diminished, when realistic noise is added.

The key question is, what kind of conclusions can be made of the observations on the basis of simulations. In the WFCAM observations of TMC-1N, the value of central density is estimated to be between $10^4$-$10^5$ cm$^{-3}$, which is within the range of our simulations. In this case, in particular method $D$ should be quite reliable according to our simulations (see Figs.~\ref{fig:p1} and ~\ref{fig:m1}). However, the simulations are simplifications of true filaments, and they do not take into account all factors that could affect the interpretations. For example, there is no background gradient in the simulations, whereas in the observed area, there is a strong gradient that could distort the obtained parameters.
A linear background is included as a free parameter in the fitting of Plummer profile, and therefore a linear background gradient perpendicular to the filament should not cause problems. However, nonlinear gradients could affect the fitting.
In the simulations the filament is also assumed to be in the plane of the sky. 
All the generated photometric errors are normally distributed. In real observations, there could be outliers, which could affect both methods $C$ and $D$.

However, the fundamental question is, whether or not the real filament profiles are really Plummer functions. In the simulations, we presume that the filament profile obeys the Plummer function, and therefore the results fit well that function unless there is some bias. In real observations, this is not necessarily the case. We have also noticed that even a poor fit can give reasonable values for the Plummer parameters. Conversely, good fits can also give, e.g., very high values for the $p$ parameter.
It is therefore advisable that all fits are checked by eye, even for large surveys.

The methods could be further improved, e.g., by carrying out a separate outlier filtering step (on $A_V$ map) before using method $D$, or weighting the profiles according to the number of stars in the median bins in method $C$. Background structure could also be modelled and subtracted before extracting the profiles. In principle, a linear background should not affect the resulting profiles, but in practise the fits seemed to be quite sensitive to any small variations in the data.

We have shown that filament profiles can be derived from NIR extinction data. These methods can serve as a valuable point of comparison to the results obtained with \emph{Herschel} or other instruments using submillimetre dust emission. The methods can be used with, e.g., UKIDSS and future NIR surveys, but we have also shown that even 2MASS data can give rather reliable estimates in the case of very nearby clouds.
NIR extinction and groundbased dust emission measurements with, e.g., SCUBA-II can be used as alternative or complementary means of studying the properties of filaments. The limits of the NIR extinction methods come from the requirements of having enough background stars. This restricts the use of these methods to low latitudes and nearby sources. The resolution of the method is limited only by the amount of background stars.

In order to achieve better resolution, NIR surface brightness could be used~\citep[see e.g.][]{Padoan2006,Juvela2006,Juvela2008}.
However, the observations of NIR surface brightness can be quite difficult. We study the properties of TMC-1N using NIR surface brightness data in a following paper (Malinen et al. 2012, in prep.).

\section{Conclusions} \label{sect:conclusions}

We have compared the filament profiles obtained by submillimetre observations (method $A$) and NIR extinction (methods $B$, $C$, and $D$).

\begin{itemize}
\item We conclude that NIR extinction maps (method $B$) can be used as an alternative method to submillimetre observations to profile molecular cloud filaments.

\item We show that if better resolution NIR data are not available, 2MASS data can also be used to estimate approximate profiles of nearby filaments.

\item We also present two new methods for the derivation of filament profiles using NIR extinction data:
\begin{itemize}
\item[$\bullet$] Plummer profile fits to median $A_{\rm V}$ values of stars within certain offset bin from the filament centre (method $C$)
\item[$\bullet$] Plummer profile fits directly to the $A_V$ values of individual stars (method $D$)
\end{itemize}

\item We show that in simulations, all the methods based on NIR extinction work best with modest densities $\rho_{\rm c}$ = $10^4$--$10^5$ cm$^{-3}$. Method $D$ gives correct results for the fitted and derived parameters to within $\sim$10\% and methods $B$ and $C$ to within $\sim$20\% in most cases.
At high densities ($\sim10^6$ cm$^{-3}$), only method $D$ continues to work reliably.

\item For the profile fits to real observations the values of individual Plummer parameters are in general similar to within a factor of $\sim$2 (in some cases up to a factor of $\sim$5). Although the Plummer parameter values can show significant variation, the derived estimates of filament mass (per unit length) usually remain accurate to within some tens of per cent.

\item We confirm the earlier results of \citet{Nutter2008} that the examined TMC-1 structure, or Bull's tail, is a filament and not part of a continuous ring. Our parameter values are also consistent with the earlier results.

\item The confidence regions of the Plummer parameters are quite wide. As the parameters are not always well constrained, even small changes in the data can lead to different parameter values. The correlations of the parameters should therefore be taken into account.

\item The size and orientation of the confidence regions are not identical when \emph{Herschel} maps or NIR data are used. The combination of the methods can significantly improve the accuracy of the filament parameter estimates.

\end{itemize}

\begin{acknowledgements}
We thank the referee, Laurent Cambr\'{e}sy, for suggestions that improved the paper.
We are grateful to Mike Irwin for reducing the WFCAM data.
JM and MJ acknowledge the support of the Academy of Finland Grants No. 250741 and 127015. JM also acknowledges a grant from Magnus Ehrnrooth Foundation. MGR gratefully acknowledges support from the Joint ALMA Observatory and the Joint Astronomy Centre, Hawaii (UKIRT).
PP is funded by the Funda\c{c}\~ao para a Ci\^encia e a Tecnologia (Portugal).
The United Kingdom Infrared Telescope is operated by the Joint Astronomy Centre on behalf of the Science and Technology Facilities Council of the U.K.
This paper makes use of WFCAM observations processed by the Cambridge
Astronomy Survey Unit (CASU) at the Institute of Astronomy, University
of Cambridge.
This publication makes use of data products from the Two Micron All Sky Survey, which is a joint project of the University of Massachusetts and the Infrared Processing and Analysis Center/California Institute of Technology, funded by the National Aeronautics and Space Administration and the National Science Foundation.
This research made use of Montage, funded by the National Aeronautics and Space Administration's Earth Science Technology Office, Computation Technologies Project, under Cooperative Agreement Number NCC5-626 between NASA and the California Institute of Technology. Montage is maintained by the NASA/IPAC Infrared Science Archive.
\end{acknowledgements}

\bibliographystyle{aa}
\bibliography{biblio_j1}

\clearpage

\Online
\begin{appendix}

\section{Comparing filament profiling in \emph{Herschel} fields and 2MASS data}  \label{sect:appendix_A}

\begin{figure*}
\includegraphics[width=16cm]{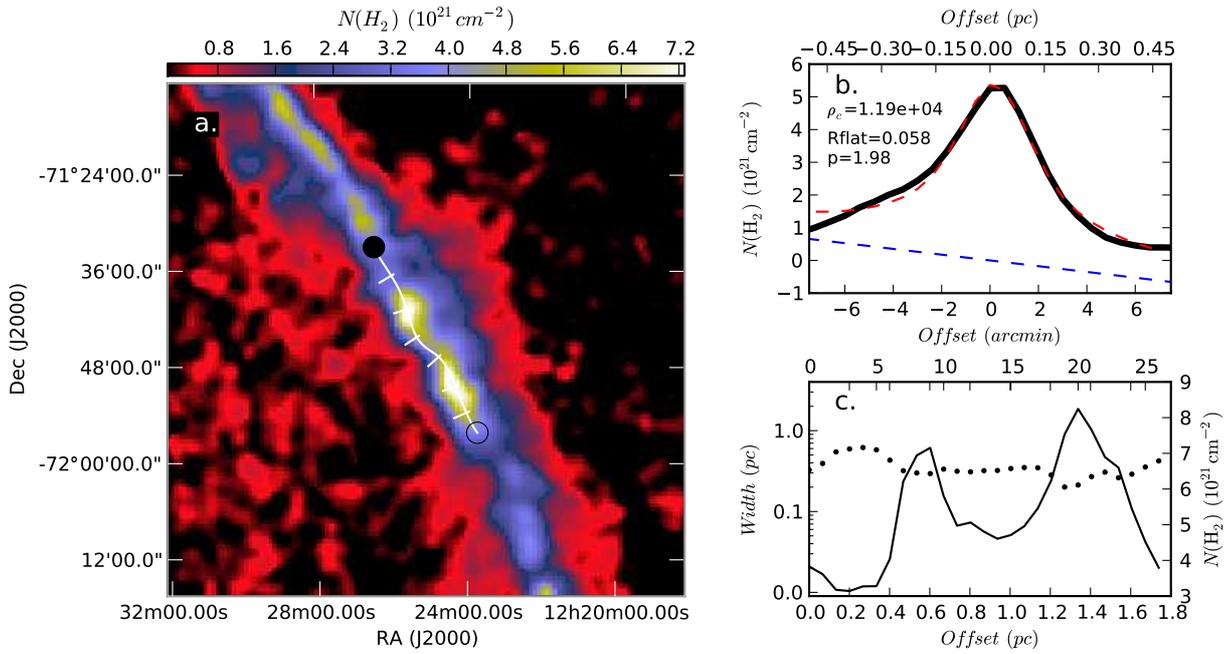}
\caption{G300.86-9.00 (PCC550) column density map and filament profile derived from 2MASS extinction map. See Fig.~\ref{fig:n_a_f} for explanations of the used notation.}
\label{fig:PCC550_maps}
\end{figure*}

\begin{figure*}
\includegraphics[width=16cm]{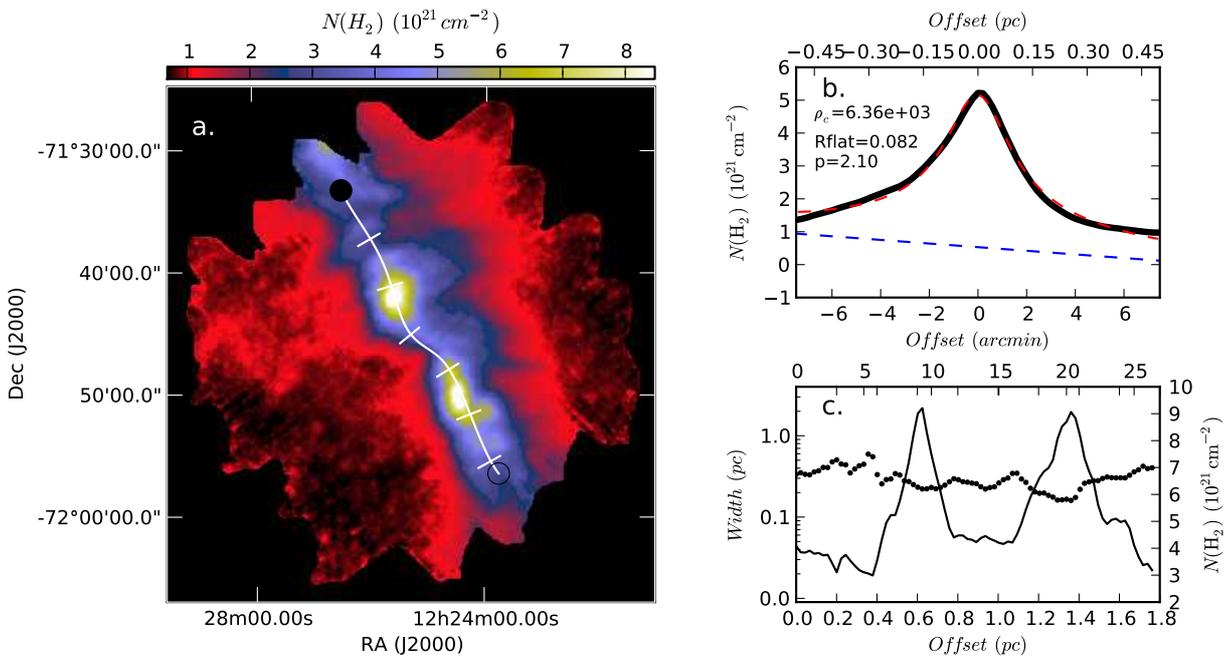}
\caption{300.86-9.00 (PCC550) column density map and filament profile derived from \emph{Herschel} emission map. See Fig.~\ref{fig:n_a_f} for explanations of the used notation.}
\label{fig:PCC550_maps_colden}
\end{figure*}

\begin{figure*}
\includegraphics[width=16cm]{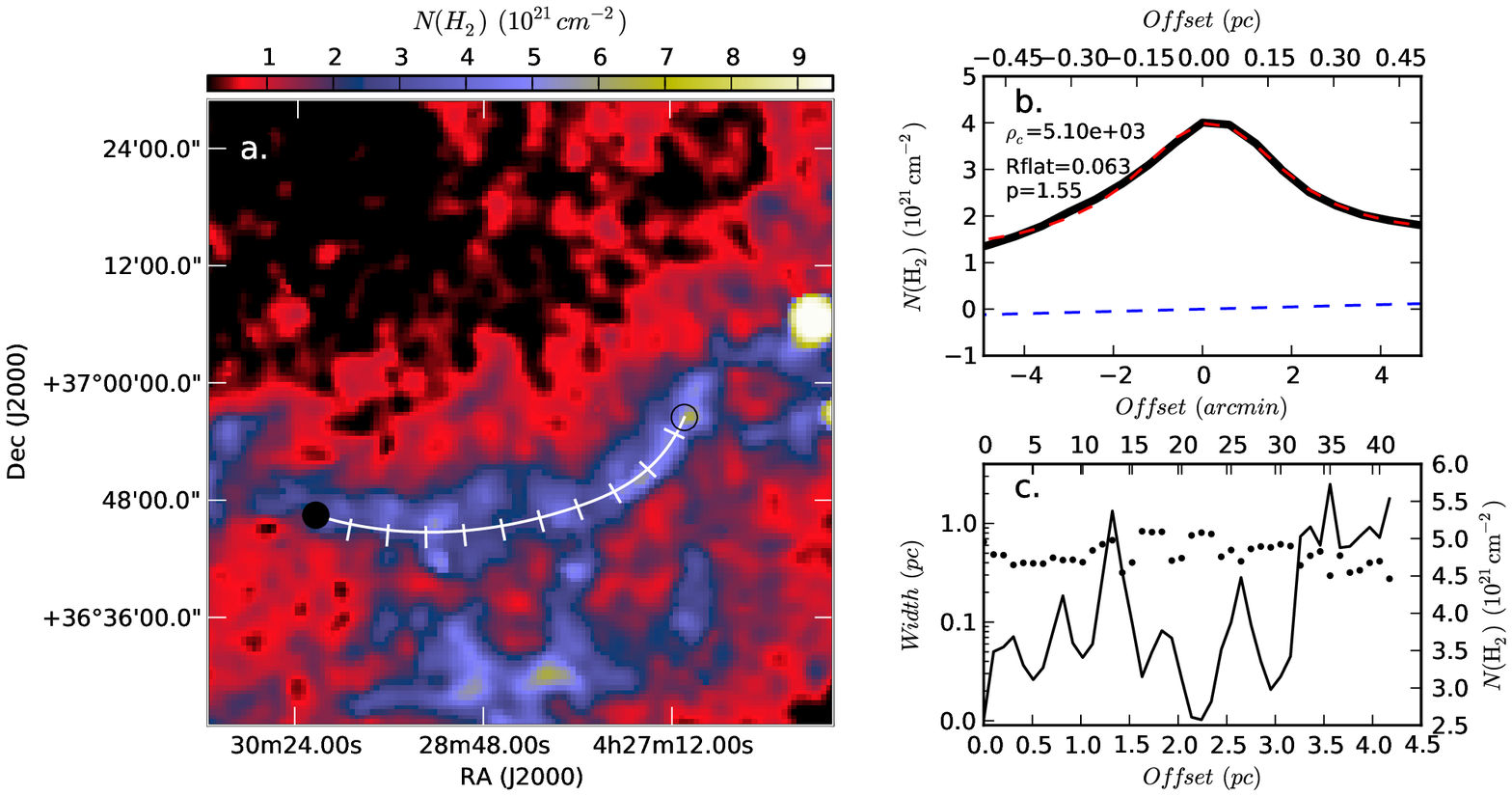}
\caption{G163.82-8.44 column density map and filament profile derived from 2MASS extinction map. See Fig.~\ref{fig:n_a_f} for explanations of the used notation.}
\label{fig:v31_7420_shift_maps}
\end{figure*}

\begin{figure*}
\includegraphics[width=16cm]{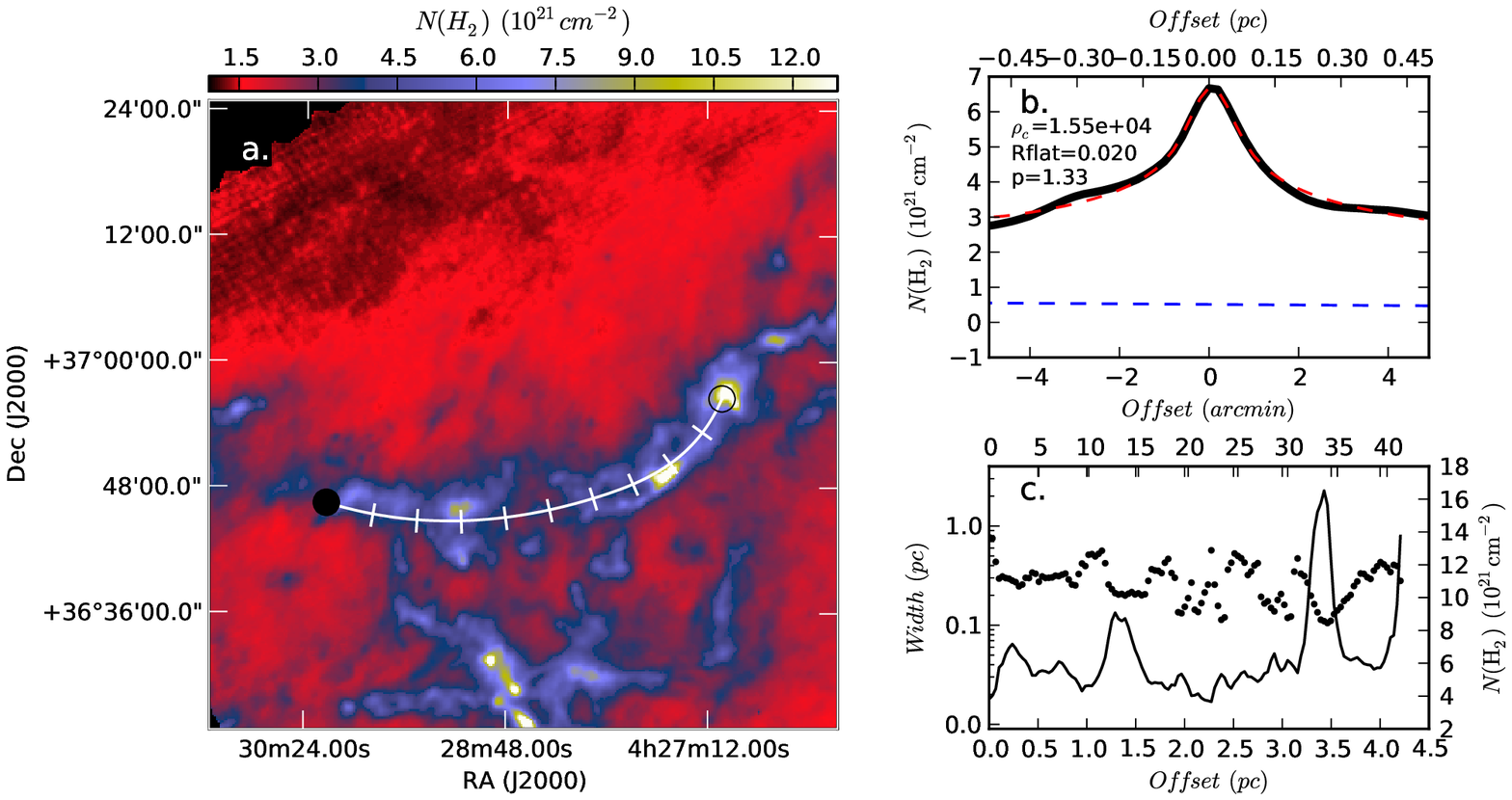}
\caption{G163.82-8.44 column density map and filament profile derived from \emph{Herschel} emission map. See Fig.~\ref{fig:n_a_f} for explanations of the used notation.}
\label{fig:v31_7420_shift_maps_colden}
\end{figure*}

We compare our methods to two profiles obtained using \emph{Herschel} column density maps presented in~\citet{Juvela2012}, namely fields G300.86-9.00 (PCC550) and G163.82-8.44. The distances to these filaments are $\sim$230\,pc and 350\,pc, respectively.
Column density maps and filament profiles derived from 2MASS extinction maps and \emph{Herschel} emission maps are shown in Figs.~\ref{fig:PCC550_maps} and ~\ref{fig:PCC550_maps_colden} for G300.86-9.00, and \ref{fig:v31_7420_shift_maps} and \ref{fig:v31_7420_shift_maps_colden} for G163.82-8.44.

\begin{figure}
\includegraphics[width=8cm]{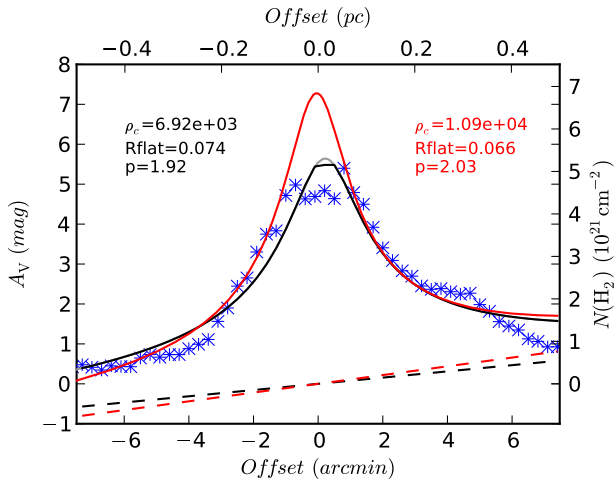}
\caption{Profiles of 300.86-9.00 (PCC550) filament derived by using methods $C$ and $D$ on 2MASS data. See Fig.~\ref{fig:stars} for explanations of the used notation.}
\label{fig:PCC550_stars}
\end{figure}

\begin{figure}
\includegraphics[width=8cm]{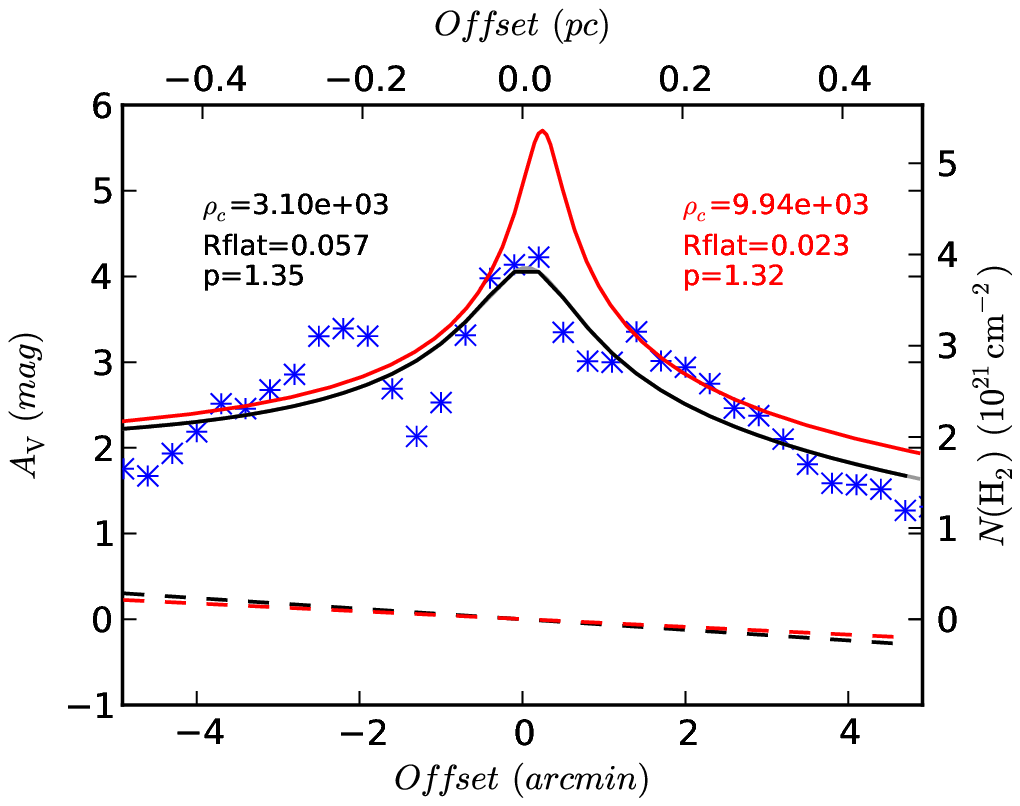}
\caption{Profiles of G163.82-8.44 filament derived by using methods $C$ and $D$ on 2MASS data. See Fig.~\ref{fig:stars} for explanations of the used notation.}
\label{fig:v31_7420_shift_stars}
\end{figure}

The filaments are traced based on the \emph{Herschel} maps and marked with white line in both column density maps. The filament positions obtained from \emph{Herschel} maps fit quite well also in the 2MASS maps. 
Profiles of the filaments derived by methods $C$ and $D$ 
using 2MASS data are shown in Figs.~\ref{fig:PCC550_stars} and \ref{fig:v31_7420_shift_stars}.
The filament parameters derived with different methods are shown in table~\ref{tab:result_param_HIII}. The differences to the results of~\citet{Juvela2012}
are caused by small modifications in the method and the change of the selected filament segment.
The distance to these filaments is greater than to TMC-1, which means that there are fewer stars in the filament area and a larger probability of foreground stars, which can bias the results.
However, in general, we obtain quite similar results using all our four methods. In these filaments anticorrelation of FWHM in the densest peaks is also quite clear in the \emph{Herschel} maps, but not in the 2MASS maps.

\begin{table*}
\caption{Obtained parameters describing the filaments G300.86-9.00 and G163.82-8.44. The columns are the data set, method (column density map ($A$), $A_{\rm V}$ map ($B$), median profile ($C$) and fit to individual stars ($D$)),
FWHM (including the effect of the data resolution), Plummer parameters and mass per unit length $M_{\rm line}$.}
\label{tab:result_param_HIII}
\begin{tabular}{lllllll}
\hline
Data set & Method & FWHM & $\rho_{\rm c}$ & $R_{\rm flat}$ & $p$ & $M_{\rm line}$\\
~                 &~       & (pc) & (cm$^{-3}$)    & (pc)           &~    & ($M_{\sun}$/pc)\\
\hline
G300.86-9.00 (PCC550) & & & & & &\\  
\hline
\emph{Herschel} maps & $A$ & 0.25 & 6.36e+03 & 0.082 & 2.10 & 3.12e+01\\
2MASS: $A_{\rm V}$ maps & $B$ & 0.31 & 1.19e+04 & 0.058 & 1.98 & 3.88e+01\\
2MASS: Median profile & $C$ & 0.31 & 6.92e+03 & 0.074 & 1.92 & 3.67e+01\\
2MASS: Fit to individual stars & $D$ & ... & 1.09e+04 & 0.066 & 2.03 & 4.10e+01\\
\hline
G163.82-8.44 & & & & & & \\  
\hline
\emph{Herschel} maps & $A$ & 0.20 & 1.55e+04 & 0.020 & 1.33 & 5.47e+01\\
2MASS: $A_{\rm V}$ maps & $B$ & 0.40 & 5.10e+03 & 0.063 & 1.55 & 4.21e+01\\
2MASS: Median profile & $C$ & 0.33 & 3.11e+03 & 0.057 & 1.35 & 3.91e+01\\
2MASS: Fit to individual stars & $D$ & ... & 9.94e+03 & 0.024 & 1.32 & 4.41e+01\\
\hline
\end{tabular}
\end{table*}

\section{Effects of YSOs}  \label{sect:appendix_B}

\begin{figure*}
\includegraphics[width=16cm]{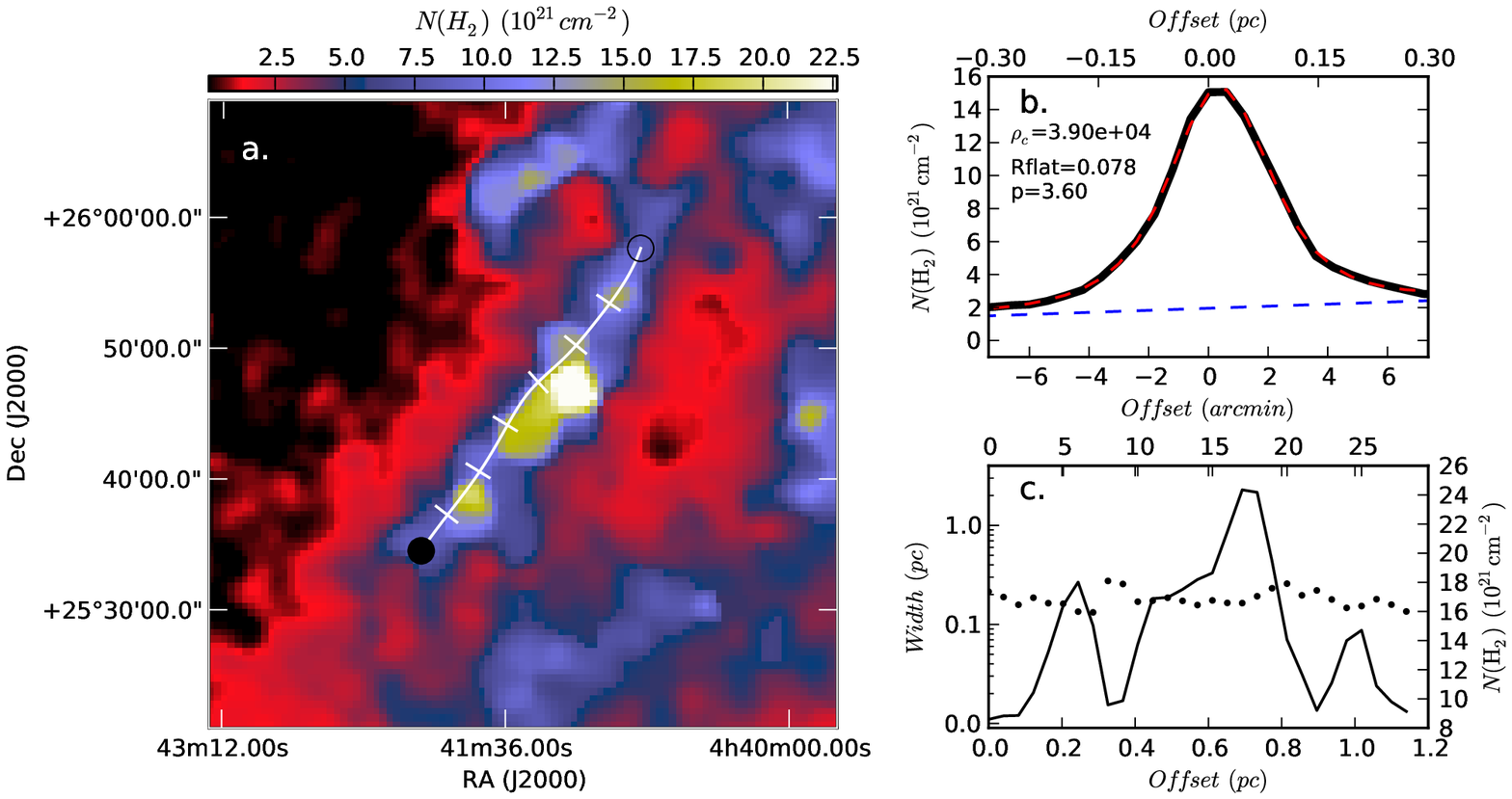}
\caption{TMC-1 column density map and filament profile derived from 2MASS extinction map (YSOs not removed). See Fig.~\ref{fig:n_a_f} for explanations of the used notation. See Fig.~\ref{fig:a_f} for a map and profiles derived after the YSOs are removed.
}
\label{fig:a_f_ysos}
\end{figure*}

YSOs may bias the extinction estimates, if they are not removed. For a comparison to Fig.~\ref{fig:a_f} made after removing the YSOs (six in the profiled filament area), we show in Fig.~\ref{fig:a_f_ysos} the column density map and profiles of TMC-1 made before removing the YSOs. Even though the derived profile parameter values for both cases are similar to within $\sim$10-30\%, notable local changes can occur in the column density maps and in the column density profiles along the ridge of the filament.

\end{appendix}

\end{document}